\documentclass[aps,prx,reprint,amsmath,amssymb,superscriptaddress, nofootinbib, 10pt]{revtex4-2}
\setcitestyle{numbers,square} 
\usepackage{xcolor}
\usepackage[dvipsnames]{xcolor}
\usepackage{empheq}

\usepackage{hyperref}
\usepackage{url}
\newcommand{\bvec}[1]{{\vec #1}}
\usepackage{physics}
\renewcommand\vec[1]{\ensuremath\boldsymbol{#1}}
\usepackage{comment}
\usepackage{graphicx}
\usepackage{caption}
\usepackage{fancyhdr}
\usepackage{ amssymb }
\usepackage{changepage} 
\usepackage{hyperref}
\usepackage{subcaption}
\usepackage[normalem]{ulem}

\usepackage{mathrsfs}
\hypersetup{
colorlinks=true,
linkcolor=blue,
filecolor=blue,
citecolor = blue,
urlcolor=blue,
}

\usepackage{graphicx}
\usepackage{dcolumn}
\usepackage{bm}

\begin{document}

\preprint{AIP/123-QED}

\title{Electron Hydrodynamics: Viscosity Tensor and effects of a Magnetic field}

\author{Anubhav Srivastava }
\author{Subroto Mukerjee}
\affiliation{ 
Department of Physics, Indian Institute of Science, Bengaluru}
\date{\today}

\begin{abstract}
Abstract: Transport due to electrons in ultra-clean two dimensional systems can be hydrodynamic in nature with the momentum of the electrons being conserved in the bulk. This hydrodynamic behavior coupled with effects of Berry curvature arising from band structure can give rise to novel vortical transport coefficients relating the stress tensor to gradients in the electrostatic potential and temperature. These coefficients have been calculated in the absence of a magnetic field and have been shown to depend only on the equilibrium distribution function~\cite{Chadha_Mukerjee2024}. In this paper, we first obtain an expression for the viscosity tensor and show that the Berry curvature generates odd components of the viscosity tensor arising from the intrinsic angular momentum of the Bloch wavepackets. We calculate the viscosity tensor for a two-dimensional microscopic model of tilted Dirac cones. We next obtain the vortical coefficients and the viscosity tensor in the presence of a magnetic field and extend the Onsager relations for them to include both the magnetic field and the Berry curvature. We show that the field dependence of the coefficients manifests itself in the non-equilibrium part of the distribution function and calculate them to second order in the electron-electron scattering time. We explicitly show that the expressions we obtain are consistent with the Onsager relations. 

\end{abstract}

\maketitle

\section{\label{sec:level1}Introduction}
In most metallic systems, electron transport is studied in the diffusive regime where momentum-relaxing collisions of electrons are dominant. However, advancements in experimental techniques have enabled the preparation of ultra-pure 2D materials, facilitating access to the hydrodynamic regime of electron transport\cite{Jong1995,Bandurin2016,Moll2016,KrishnaKumar2017,Bandurin2018,majumdar2025universalityquantumcriticalflow}. In such systems, the momentum-conserving collisions of electrons amongst themselves dominate over the momentum-relaxing collisions with phonons and impurities, which leads to the momentum density having a well-defined dynamics on long time scales. Such electron transport exhibits several interesting phenomena such as the violation of the Sharvin bound on conductance\cite{Bandurin2016}, a strong violation of Weidmann-Franz law\cite{majumdar2025universalityquantumcriticalflow} and the formation of whirlpools leading to negative-vicinity resistance\cite{Danz2020,Polini2020,Narozhny2022a,Narozhny2022b,Palm2024}.

In the hydrodynamic description, only conserved quantities exhibit well-defined dynamics at time and length scales much larger than the microscopic scales. Thus, in the high-purity limit, with negligible momentum-relaxing collisions, electron transport exhibits hydrodynamic behaviour characterized by the conservation of three quantities \textit{viz} charge, entropy, and momentum densities. These conserved modes dominate the long-wavelength, low-frequency dynamics, enabling a fluid-like description of electron flow. Assuming weak, slowly varying perturbations, the system's response can be expressed linearly in terms of these perturbations\cite{Chaikin1995}. 

Unlike in magneto-hydrodynamic systems, the dynamics of an electron in a crystal can be influenced by the non-trivial winding of the periodic part of Bloch wavefunctions in the Brillouin zone. The winding is encoded in the gauge-invariant Berry Curvature(or the geometric curvature denoted by $\boldsymbol{\Omega}$) that acts as a magnetic field in momentum space. The semi-classical approach of electron dynamics can be applied to model electron transport under weak perturbations with inter-band transitions suppressed. A wave packet is constructed as a linear superposition of Bloch states such that 
$\Delta k \ll \frac{1}{a} \text{ and } \Delta x \gg a$, (where $\vec{k}$ is the wavevector, $\vec{x}$ is the position and $a$ is the size of the unit cell) which then evolves according to the semi-classical Ehrenfest equations of motion. The Berry curvature modifies the equation for $\dot{\boldsymbol{x}}$ through the addition of an anomalous term $\dot{\boldsymbol{k}}\times \boldsymbol{\Omega}$. The effects of geometry on transport properties can be studied using Boltzmann transport theory and semiclassical dynamics, providing a unified framework to explain a wide range of interesting phenomena such as the quantum anomalous Hall effect, valley Hall effect, and the Nernst Hall effect\cite{Girvin_Yang_2019,Xiao2010}.

An important transport coefficient in hydrodynamic systems is the fourth-rank viscosity tensor($\eta_{ij,kl}$) relating the stress tensor to the strain rate. In systems where the constituent particles possess no internal rotational degrees of freedom, the conservation of external angular momentum demands that the stress tensor be symmetric. However, the Berry curvature induces an orbital magnetization for the electrons, which can be physically understood to arise from the rotation of the wavepacket about its centre\cite{Panigrahi_Mukerjee2023}. This leads to the stress tensor being asymmetric. The breaking of time-reversal symmetry can cause the viscosity tensor to acquire dissipation-less components called odd viscosity components in addition to the usual shear and bulk viscosity components.

In the hydrodynamic regime, velocity gradients can lead to charge and heat currents, while temperature and potential gradients can, in turn, cause a flux of momentum. The geometric curvature causes the associated transport coefficients to be vortical\cite{Chadha_Mukerjee2024}. Onsager's relations arising from microscopic reciprocity of time are relations between components of the transport coefficients. In addition to Onsager's relations, the cross-tensor coefficients also obey Curie's symmetry principle, \textit{i.e.}, they vanish in centrosymmetric systems\cite{GrootBook2013}. The total thermodynamic currents contain a magnetization contribution, which does not contribute to transport. Onsager's relations apply only to the coefficients of transport currents obtained by subtracting magnetization components and not to the total currents\cite{Cooper1997,Xiao2006Anomalous}. The total currents comprise contributions from equilibrium and out-of-equilibrium parts of the distribution. Ref. \cite{Chadha_Mukerjee2024} calculates the cross-tensor transport coefficients from the equilibrium part of the distribution function and shows that Onsager's relations and Curie's principle remain valid. However, to account for the effects of the magnetic field, we need to solve for the out-of-equilibrium part of the distribution function\cite{Ziman1972,Panigrahi_Mukerjee2023}.

The first result of this paper is the derivation of the anomalous viscosity tensor from the stress-magnetization tensor. The viscosity tensor calculated therein does not contribute to dissipation and hence only has odd components. It is analogous to the anomalous Hall conductivity tensor of Chern insulators and is not forbidden to exist by isotropy in 2D systems. A time-reversal breaking Berry curvature can, in principle, cause the odd viscosity to become finite\cite{Avron1998}.  We explicitly show that the stress tensor need not be symmetric due to the contribution of the orbital magnetic moment of electrons to the total angular momentum, thereby causing the viscosity to have non-zero odd components. Finally, we use the expression we obtain to numerically calculate the odd viscosity components for a microscopic model of tilted Dirac cones \cite{SodemmanFu}.
In the second part of the paper, we employ the semi-classical approach coupled with Boltzmann transport theory to obtain the non-equilibrium part of the distribution function. We calculate the cross-tensor transport coefficients to second order in the electron-electron scattering time($\tau_{ee}$) and demonstrate that Onsager's reciprocity relations and the Curie symmetry principle continue to hold. We show that in the most general case, vortical currents can arise from a magnetic field. In the course of the paper, we point out an important aspect of Onsager's relations in systems with a finite Berry Curvature, namely, if the time-reversal symmetry is broken, the Berry curvature must be transformed $\boldsymbol{\Omega}_{\vec{k}}\rightarrow -\boldsymbol{\Omega}_{-\vec{k}}$  on the other side of the equation in Onsager's relations. 
\section{Onsager's reciprocity relations}
Onsager's reciprocity relations are a set of relations between transport coefficients and are based on microscopic reversibility in time. The relations are closely related to the trajectory traced by the particle if its momentum coordinates are reversed. In a 2D system with Berry Curvature, the decoupled equations of motion are \cite{Xiao2010},
\begin{subequations}
\label{eq_EOM2D}
\begin{align} 
\dot{\bvec{x}} &=\frac{
\frac{1}{\hbar} \vec{\nabla_k}\varepsilon_{\vec{k}} 
+ \frac{e}{\hbar} (\vec{E} \times \vec{\Omega_k}) 
}{ 1 + \frac{e}{\hbar} \vec{B} \cdot \vec{\Omega_k} }
\\  \dot{\vec{k}} & =
-\frac{\frac{e}{\hbar} \vec{E} 
+ \frac{e}{\hbar^2} \vec{\nabla_k}\varepsilon_{\vec{k}} \times \vec{B} 
}{ 1 + \frac{e}{\hbar} \vec{B} \cdot \vec{\Omega_k} }
\end{align}
\end{subequations}
When the momentum($\vec{k}$) of the particle is reversed it retraces its path provided the following transformations are also made: $\vec{B}\rightarrow -\vec{B}$, $\vec{\Omega_{\vec{k}}}\rightarrow -\vec{\Omega_{-\vec{k}}}$  and $\varepsilon_{\vec{k}}\rightarrow \varepsilon_{-\vec{k}}$. This fact is of significance in the derivation of Onsager's relations\cite{Luo2020}.

Since momentum is conserved during interparticle collisions, it remains a well-defined quantity in hydrodynamic systems. The corresponding flux, represented by the stress tensor and its conjugate statistical field, the velocity field are well-defined quantities. Consequently, velocity gradients can, in principle, drive electric and heat currents while gradients in electric potential and temperature can drive momentum currents. In the linear response regime, we can write the fluxes to linear order in the gradients, thus defining the transport coefficients\cite{Groot1954,Chadha_Mukerjee2024}.
\begin{subequations}
\label{Transport_eq}
\begin{align}
j^N_i &= -L^1_{i,j} \partial_j \phi - L^2_{i,j} \partial_j T - L^3_{i,jk} \partial_k u_j \label{eq_linear_charge} \\
j^Q_i &= -L^4_{i,j} \partial_j \phi - L^5_{i,j} \partial_j T - L^6_{i,jk} \partial_k u_j \label{eq_linear_heat} \\
\pi_{i,j} &= -L^7_{ij,k} \partial_k \phi - L^8_{ij,k} \partial_k T - L^9_{ij,kl} \partial_l u_k, \label{eq_linear_Stress}
\end{align}
\end{subequations}
where $\phi$ is the electric potential, $T$ is the temperature and $\vec{u}$ is the velocity field. The subscripts indicate spatial indices. It can be seen that while the charge and heat currents($j^N_i$ and $j^Q_i$ respectively) are vectors, the momentum flux (or the stress tensor denoted by $\pi_{i,j}$) is a second-rank tensor. The transport coefficients $ L^3$ and $ L^6 $ relate a second-rank tensor(velocity field gradient) to a vector quantity, the charge current and heat current, respectively. Hence, $ L^3 $ and $ L^6 $, and by a similar reasoning $ L^7 $ and $ L^8 $, are classified as cross-tensor transport coefficients. According to Curie's symmetry principle, cross-tensor transport coefficients vanish in centrosymmetric systems~\cite{GrootBook2013}.

Microscopic reversibility along with the Boltzmann entropy postulate can be used to derive relations between transport coefficients. The resulting Onsager's relations(see Appendix \hyperref[Appendix_Onsager_der]{A1}) are:
\begin{subequations}
\label{eq_Onsager_relatios}
\begin{align}
 \label{eq_Onsager_relatiosA}L^3_{i,jk} (\vec{B};\{\vec{\Omega_{k}}\}, \{\varepsilon_{\vec{k}}\} ) &= -L^7_{jk,i}(-\vec{B};-\{\vec{\Omega_{-k}}\}, \{\varepsilon_{\vec{-k}}\} ) 
  \\ \label{eq_Onsager_relatiosB}L^6_{k ,ji}(\vec{B};\{\vec{\Omega_{k}}\}, \{\varepsilon_{\vec{k}}\} ) 
 &=-T L^8_{ji,k}(-\vec{B};-\{\vec{\Omega_{-k}}\}, \{\varepsilon_{\vec{-k}}\} )\\
 \label{eq_Onsager_relatiosC}
 L^9_{ij,kl}(\vec{B};\{\vec{\Omega_k}\}, \{\varepsilon_{\vec{k}}\} )&=L^9_{kl,ij}(-\vec{B};-\{\vec{\Omega_{-k}}\}, \{\varepsilon_{\vec{-k}}\} ) 
\end{align}
\end{subequations}
The curly brackets in the above equation denote that the transport coefficients are a functional of the Berry curvature and band dispersion. The relations can be extended to their local form, where the transport coefficients are position-dependent.
\section{The Anomalous Viscosity Tensor}
\subsection*{Stress Magnetization tensor}
Typical transport measurements do not measure the total current, as the latter includes contributions from magnetization currents, which do not contribute to net transport. These magnetization currents are not gauge-invariant and give rise to circulating currents within the interior of the system. This can be seen by observing the continuity relations, which are a consequence of conservation laws\cite{Chadha_Mukerjee2024}, \begin{subequations}
\begin{align}
-e \partial_t \bar{n} + \vec{\nabla} \cdot \vec{j}^N &= 0 \label{eq_continuity_charge} \\
\partial_t \bar{\epsilon} + \vec{\nabla} \cdot \vec{j}^E &= 0, \label{eq_continuity_heat} \\
\partial_t \bar{g}_i + \nabla_j \pi_{i,j} &= -F_i \label{eq_continuityc},
\end{align}
\end{subequations}
where is $\bar{\epsilon}$ is the energy density, $\bar{\vec{g}}$ the momentum density, and we have separated all external forces($\vec{F}$). It can be seen that on adding a divergence-less quantity to the total currents, the transport is unaffected. In equilibrium, these magnetization currents can be written as the curl of the relevant magnetization densities\cite{Xiao2006Anomalous}. Specifically for the stress magnetization current, we can write \cite {Chadha_Mukerjee2024,Cooper1997},
\begin{equation}
[\pi_{M}]_{ij}=\epsilon_{jpl}\pdv{[M_0^{\pi}]_{il}}{x_p}+ \epsilon_{jpl} \pdv{\psi}{x_p}[M^{\pi}_0]_{il},
\label{eq_stress_tensor_mag_form}
\end{equation}
Here $\psi$ is the gravitational potential, the external field that couples to energy density, and $M_0^{\pi}$ is the equilibrium stress magnetization tensor.  Thus, to obtain the physically observable transport currents the magnetization currents must be subtracted from total currents. The resulting transport currents must vanish in equilibrium and thus, Onsager's reciprocity and the Einstein relations hold only for the coefficients obtained from transport currents\cite{Cooper1997, Chadha_Mukerjee2024,Panigrahi_Mukerjee2023}.  The magnetization densities can be calculated by demanding that the Einstein relations be satisfied for transport currents. Following this approach, Ref.~\cite{Chadha_Mukerjee2024} calculates the equilibrium stress magnetization tensor to be
\begin{equation}
    \left({M}_0^\pi\right)_{ij}=-\frac{1}{\beta}\int{[d\vec{k}]k_i[\Omega_{\vec{k}}]_j\log(1+e^{-\beta(\epsilon_{\vec{k}}-\mu-\hbar\vec{u}.\vec{k})})}
    \label{eq_Magnetization_tensor}
\end{equation} 

 \subsection*{The Viscosity Tensor}
In fluids with constituent particles without any internal rotational degree of freedom, conservation of angular momentum constrains the stress tensor to be symmetric. Moreover, if we assume that the free energy of the system is invariant under rigid rotations, $L^9_{ij,kl}$, the viscosity tensor can be taken to be symmetric under the exchange $i\leftrightarrow j$ and  $k\leftrightarrow l$\cite{Chaikin1995}. The contribution to entropy production from components of the viscosity tensor for which the ordered pair $\{i,j\}\neq\{k,l\}$ vanishes if they are antisymmetric under the exchange $\{i,j\}\leftrightarrow\{k,l\}$. These components are called odd viscosity components and do not cause any dissipation\cite{Fruchart2023}. However, in the electron fluid, the wavepacket formed by linearly superposing Bloch waves can rotate about its centre and contribute to internal angular momentum \cite{Xiao2010}. This implies that the stress tensor need not be symmetric because an infinitesimal volume element might exchange internal and external angular momentum\cite{GrootBook2013}.

The stress tensor can be obtained from the stress-magnetization density tensor employing equations \ref{eq_stress_tensor_mag_form} and \ref{eq_Magnetization_tensor}.
Setting  $\psi=0$ in equation \ref{eq_stress_tensor_mag_form}, and extracting the coefficient of $\nabla \vec{u}$ according to equation \ref{eq_linear_Stress},  we obtain 
\begin{equation} 
L^9_{ij,mn}=-\epsilon_{jnl} \int [d\vec{k}] k_i k_m [\vec{\Omega}_{\vec{k}}]_lf^0( \vec{x},\vec{k})
\label{eq_L9_visc}\end{equation}
 Assuming isotropy of dispersion and Berry curvature, $ \varepsilon_{\vec{k}} = \varepsilon_{|\vec{k}|} \text{ and } \vec{\Omega}_{\vec{k}} = \boldsymbol{\Omega}_{|\vec{k}|}$, only components with $j\neq n$ and $i=m$ survive. Therefore the finite components are,\begin{align}\nonumber     
L^9_{xx,xy}=L^9_{yx,yy}=-L^9_{xy,xx}=-L^9_{yy,yx}=\\-\int [d\vec{k}] k_x^2  [\vec{\Omega}_{\vec{k}}]_zf^0( \vec{x},\vec{k}) \label{eq_isotropic_viscous_tensor}
\end{align}
As shown by Equation~\ref{eq_L9_visc}, the viscosity tensor is not symmetric under the exchange \( i \leftrightarrow j \) and \( m \leftrightarrow n \), reflecting the contribution of rotating wavepackets to total angular momentum.  Equation~\ref{eq_isotropic_viscous_tensor} demonstrates that dissipationless odd viscosity components remain finite at \( 0\,\mathrm{K} \), even when the chemical potential lies in the band gap, while conventional shear and bulk viscosity components vanish.  Thus, odd viscosity components can be referred to as anomalous viscosity tensor components in analogy with the anomalous Hall conductance for electrical conductivity. Odd viscosity has been reported in the context of quantum hall fluids\cite{AvronQuantumHall,ReadHallVisc,TransportSign} and electronic systems\cite{Berdyugin2019,Xian2023,messica2025stokesflowelectronicfluid}. However, to the best of our knowledge, ours is the first derivation of the odd viscosity components in terms of the Berry curvature arising from band structure.
\begin{widetext}

\subsection*{The anomalous viscosity tensor for a microscopic model}\label{subsec_microscopic}
\begin{figure}[h!]
\centering
\begin{subfigure}{.5\textwidth}
  \centering
  \includegraphics[width=1\linewidth]{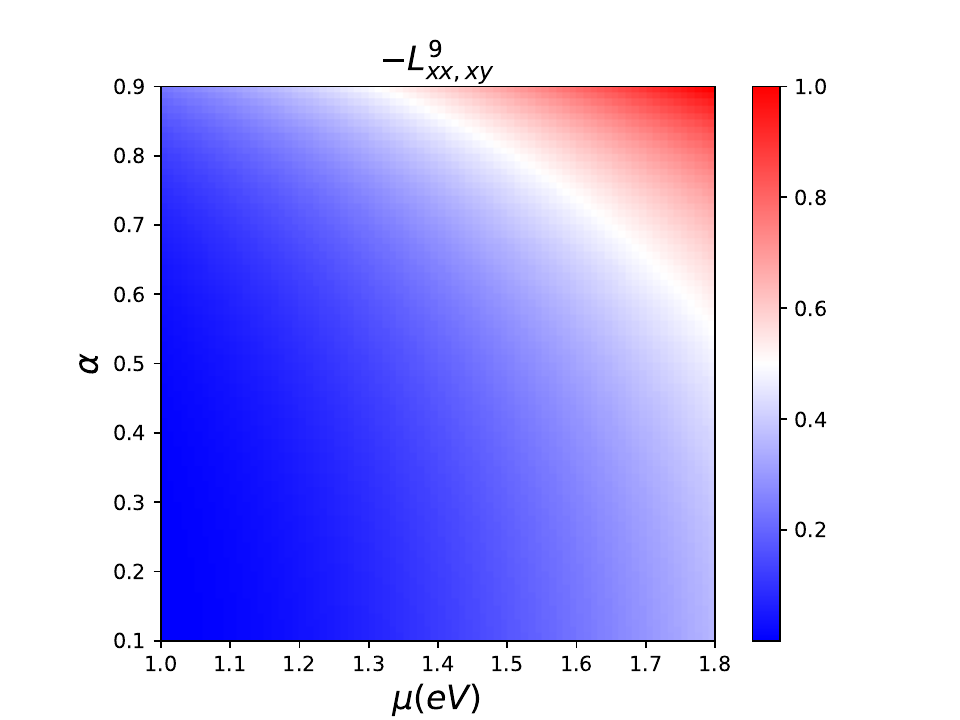}
  \caption{$L^9_{xx,xy}$ for $s=1$}
  \label{fig7.2sub1}
\end{subfigure}%
\begin{subfigure}{.5\textwidth}
  \centering
  \includegraphics[width=1\linewidth]{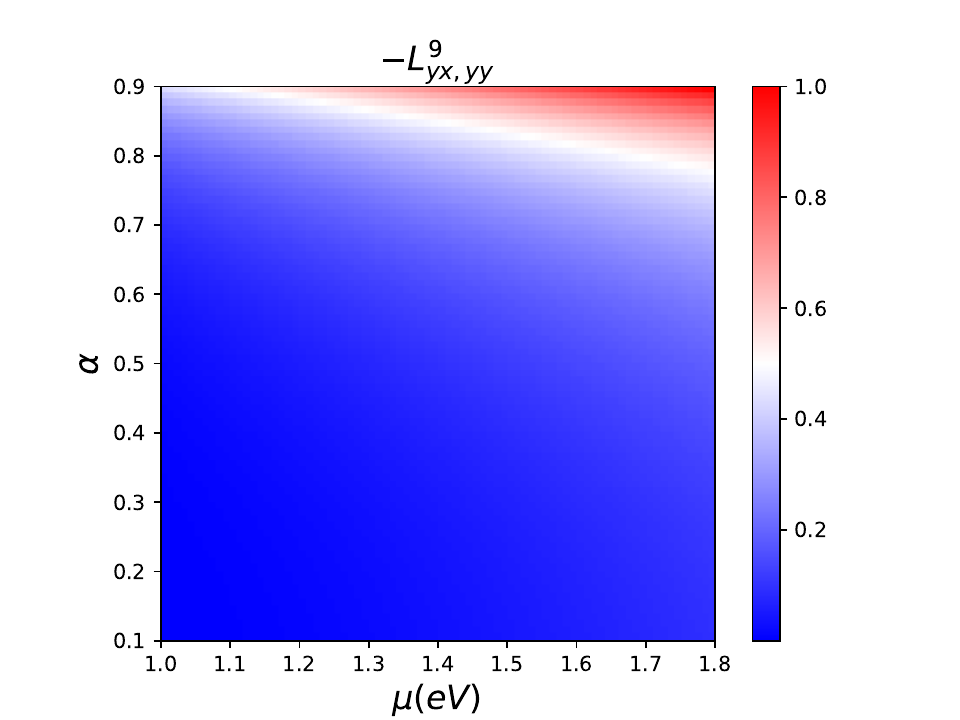}
  \caption{$L^9_{yx,yy}$ for $s=1$}
  \label{fig7.2sub2}
\end{subfigure}
\caption{The non-zero independent components of $L^9$ for the model in eq. \ref{eq_Ham}. The data set has been rescaled such that the maximum value is 1.  We have set $\Delta=1 eV$ and $k_B T= 8.167\cdot 10^{-3} eV$ absorb $v$ in the variable of integration. }
\label{fig_microscopic}
\end{figure}
    
\end{widetext}
Now, we calculate the Viscosity tensor for a microscopic model given by titled Dirac cones\cite{SodemmanFu,Chadha_Mukerjee2024}

\begin{equation}       
       H(k_x,k_y)_s = vk_x\sigma_y - svk_y \left(\sigma_x - \alpha \right) + \Delta \sigma_z
     \label{eq_Ham}
\end{equation}

The term $\alpha$ produces a tilt in Dirac cones. We have two valleys, labeled by $s$, but not separated in $\vec{k}$ space. The system breaks inversion symmetry. $\Delta$ is the onsite potential which opens up a band gap in the spectrum of both the valleys, and  $v$ is the dispersion velocity when $\Delta=0$.
The energy dispersion for the two valleys and Berry curvature has been calculated in Refs. \cite{Chadha_Mukerjee2024} and \cite{SodemmanFu}. We use $+$ and $-$ signs to denote the two distinct bands in the system.
\begin{equation}
    \epsilon_{s\pm}({\vec{ k}}) = \alpha
    s  v k_y \pm \sqrt{\left(v^2k^2 + \Delta^2\right)}
    \label{eq_dispersion}
\end{equation}
 The Berry curvature is given by
\begin{equation} 
    \Omega_{s\pm} ({\vec{k}}) =\pm \frac{1}{2} \frac{sv^2\Delta}{\left(v^2k^2 + \Delta^2\right)^{3/2}}
    \label{eq_Berry_curvature}
\end{equation}
Using equation \ref{eq_L9_visc} and summing over both the valleys, we calculate the $L^9$ from the $+$ band.

\begin{align}\nonumber
L^9_{xx,xy}&=-\sum_{s} \int [d\vec{k}] k_x ^2 \Omega_{s+} (k)f^0( \vec{x},\vec{k}) \text{ and,  }  \\ &L^9_{yx,yy}=-\sum_{s} \int [d\vec{k}] k_y ^2 \Omega_{s+} (k)f^0( \vec{x},\vec{k})
\label{eq_L9_visc_model}\end{align}
Since the Berry-Curvature has opposite signs for both valleys and is only dependent on $|\vec{k}|$, the contribution from one valley would exactly cancel the contribution from the other valley(this can be seen by a change in the variable of integration $k_y\rightarrow-k_y$). We plot the contribution from one of the valleys. Physically this can be realized by raising the energy of one of the valleys such that only the other one is filled. This amounts to time-reversal symmetry breaking in the system.

Figure \ref{fig_microscopic} shows the non-zero independent components of $L^9$ tensor. Since the dispersion is not isotropic, $L^9_{xx,xy}\neq L^9_{yx,yy}$. We see that as the tilt $\alpha$ decreases, the $L^9$ tensor decreases; however, unlike the vortical coefficients, it does not approach 0 \cite{Chadha_Mukerjee2024}. Additionally, we observe that $L^9$ increases as the chemical potential is increased, which is expected because more states contribute to the integral.

\section{The Out of Equilibrium Distribution function}
Now, we calculate the out-of-equilibrium distribution function, which is essential for a complete characterization of transport coefficients beyond the strong correlation limit (\(\tau_{ee} \rightarrow 0\))~\cite{Toshio2020,Hasdeo2021}. We adopt a semiclassical approach, assuming that inter-band transitions occur at an energy scale \(\Delta\) that is significantly larger than both the thermal energy and the energy scale associated with perturbing fields, restricting the dynamics to a single band. In the presence of electric and magnetic fields, the heuristic criteria are \(eEa \ll \Delta^2 / \varepsilon_F\) and \(\hbar eB / m_e \ll \Delta^2 / \varepsilon_F\)~\cite{Girvin_Yang_2019}. In the strong correlation limit, the frequent electron-electron scattering causes rapid equilibration of electrons, causing the system to be in local thermal equilibrium. Away from this limit, the total distribution function is written as the sum of two components, a local thermal equilibrium distribution \(f^{\text{lab}}\), and an out-of-equilibrium correction \(g\). We assume that the out-of-equilibrium correction is much smaller than the equilibrium component, i.e., \(g / f^{\text{lab}} \ll 1\).

The equilibrium part of the distribution function can be found by boosting to the frame where the fluid is locally at rest. The Doppler-shifted energy of the wave packet in this frame is $\varepsilon_{\vec{k}}-\hbar\vec{u}( \vec{x})\cdot \vec{k}$ (where $\vec{u}( \vec{x})$ is the velocity profile of fluid in lab frame). Since the fluid is locally in thermal equilibrium in this frame we can write the distribution function in the frame of the fluid as\cite{Chaikin1995}, 
\begin{equation}
    f^{\text{fluid}}( \vec{x}, \vec{k}) = \frac{1}{e^{\beta(\varepsilon_{\vec{k}}^{\text{fluid}} - \mu( \vec{x}))} + 1}
    \label{eq_dist_function_fluid}
\end{equation}
In the lab frame this can be written by substituting for the Doppler shifted energy, 
\begin{equation}
    f^{\text{lab}}( \vec{x}, \vec{k}) = \frac{1}{e^{\beta(\varepsilon_{\vec{k}} - \mu( \vec{x}) - \hbar\vec{u}( \vec{x}) \cdot \vec{k})} + 1}
    \label{eq_dist_function_lab}
\end{equation}An implicit assumption in writing the above formula is that the velocity field $\vec{u}( \vec{x})$ varies on length scales much larger than the range of interparticle interactions or the characteristic length scale associated with variation in any external perturbation. Next, we employ the Boltzmann Transport formalism using the relaxation-time approximation to model electron-electron collisions to solve for the out-of-equilibrium distribution function. The Boltzmann Transport Equation which is based on Liouville's theorem is given by\cite{Girvin_Yang_2019}
\begin{equation}
   \dot{ \vec{x}}\cdot\pdv{f( \vec{x},\vec{k})}{ \vec{x}}+\dot{\vec{k}}\cdot\pdv{f( \vec{x},\vec{k})}{\vec{k}}= \left(\pdv{f( \vec{x},\vec{k})}{t}\right)_{coll}
   \label{eq_Boltzmann}
\end{equation}
However, the equations of motion in the presence of Berry curvature conserve the modified phase space volume $\sim\Delta V D_{\vec{k}}=\Delta V\left(1+\frac{e}{\hbar}\vec{B}\cdot\vec{\Omega_k}\right)$\cite{Di2005,Xian2023}. Thus the metric must be multiplied by $ D_{\vec{k}}$ when calculating fluxes. Under the relaxation time model or Bhatnagar-Gross-Krook formalism, we have $\left(\dv{f}{t}\right)_{coll}=-\frac{g( \vec{x},\vec{k})}{\tau_{ee}}=-\frac{f( \vec{x},\vec{k})-f^{\text{lab}}( \vec{x},\vec{k})}{\tau_{ee}}$, where $g( \vec{x},\vec{k})$ is the out-of-equilibrium part of distribution function\cite{Bhatnagar1954,Girvin_Yang_2019,Panigrahi_Mukerjee2023}. Substituting for $g( \vec{x},\vec{k})$ in eq. \ref{eq_Boltzmann}, we get
\begin{equation}
   \frac{g( \vec{x},\vec{k})}{\tau_{ee}}+\dot{ \vec{x}}\cdot\pdv{g( \vec{x},\vec{k})}{ \vec{x}}+\dot{\vec{k}}\cdot\pdv{g( \vec{x},\vec{k})}{\vec{k}}= -\dot{ \vec{x}}\cdot\pdv{f^{\text{lab}}( \vec{x},\vec{k})}{ \vec{x}}-\dot{\vec{k}}\cdot\pdv{f^{\text{lab}}( \vec{x},\vec{k})}{\vec{k}}
   \label{eq_Boltzmann_simplified1}
\end{equation}

From the equations of motion(eq. \ref{eq_EOM2D}), we substitute for $\dot{ \vec{x}}$ and $\dot{\vec{k}}$ on the RHS of eq. \ref{eq_Boltzmann_simplified1} and retain the terms up to first order in the perturbation(see Appendix \hyperref[Appendix_RHS_Boltz]{A2}). The equation simplifies to,
\begin{equation}
   \frac{g( \vec{x},\vec{k})}{\tau_{ee}}+\dot{ \vec{x}}\cdot\pdv{g( \vec{x},\vec{k})}{ \vec{x}}+\dot{\vec{k}}\cdot\pdv{g( \vec{x},\vec{k})}{\vec{k}}=  \frac{1}{\hbar}\frac{\partial{f^{\text{lab}}( \vec{x},\vec{k})}}{{\partial\varepsilon_{\vec{k}} }}\frac{\nabla_{\vec{k}} \varepsilon_{\vec{k}}}{D_{k} }\cdot \vec{S},
   \label{eq_Boltzmann_simplified2}
\end{equation}
where we have defined \begin{equation}
\vec{S}=e\vec{E}+e \vec{u}({ \vec{x}})\times \vec{B} +\nabla_{ \vec{x}} \mu{( \vec{x})} +\frac{\nabla_{ \vec{x}} T{( \vec{x})}}{T( \vec{x})}\bigl(\varepsilon_{\vec{k}}-\mu( \vec{x}))+\hbar \bigl(\nabla_{ \vec{x}} \vec{u}({ \vec{x}}) \bigr)\vec{k}
\label{eq_S_2d}
\end{equation}
Under conditions of weak magnetic field($\omega =\frac{eB}{m}\ll \frac{1}{\tau_{ee}}$), we can solve for $g(\vec{x},\vec{k})$  up to different orders in $\tau_{ee}$(see Appendix \hyperref[Appendix_out_of_eq]{A3}). The first-order solution is given by
\begin{equation}
g_1( \vec{x}, \vec{k}) =\frac{ \tau_{ee} }{\hbar}\frac{\partial f^{\text{lab}}( \vec{x}, \vec{k})}{\partial \varepsilon_{\vec{k}}} \frac{\nabla_{\vec{k}} \varepsilon_{\vec{k}}}{D_{\vec{k}}} \cdot \vec{S}
\label{eq_g1_first_order_2d}
\end{equation}
Substituting $g_1( \vec{x},\vec{k})$ back into equation eq. \ref{eq_Boltzmann_simplified2}, we obtain the second-order dependence of the out-of-equilibrium distribution function on $\tau_{ee}$, \begin{widetext}
\begin{align}
  \nonumber 
 g_2&( \vec{x},\vec{k}) =\frac{e\tau_{ee}}{\hbar^3D_{\vec{k}}}\left(\left(\nabla_{\vec{k}}\varepsilon_{\vec{k}}\times \vec{B}\right)\cdot\frac{\partial \left(\frac{\tau_{ee}}{D_{\vec{k}}}\frac{\partial{f^{\text{lab}}( \vec{x},\vec{k})}}{{\partial\varepsilon_{\vec{k}} }}\right) }{\partial \vec{k}}\right)(\nabla_{\vec{k}}\varepsilon_{\vec{k}}\cdot \vec{S})  \nonumber +\\ & \frac{e\tau_{ee}^2}{\hbar^3D_{\vec{k}}^2}\left(\frac{\partial{f^{\text{lab}}( \vec{x},\vec{k})}}{{\partial\varepsilon_{\vec{k}} }}\right)\bigl(\left(\nabla_{\vec{k}}\varepsilon_{\vec{k}}\times \vec{B}\right)\cdot((\vec{S}\cdot\nabla_{\vec{k}} )\nabla_{\vec{k}}\varepsilon_{\vec{k}})\bigr)
    +\frac{e\tau_{ee}^2}{\hbar^2D_{\vec{k}}^2}\frac{\partial{f^{\text{lab}}( \vec{x},\vec{k})}}{{\partial\varepsilon_{\vec{k}} }}\epsilon_{imn}\frac{\partial\varepsilon_{\vec{k}}}{\partial k_m}\frac{\partial \varepsilon_{\vec{k}}}{\partial k_p} B^n\frac{\partial u( \vec{x})^i}{\partial x^p} \label{eq_g2_second_order_2d_eq_Main} \end{align}
\end{widetext}   
In equations \ref{eq_g1_first_order_2d} and \ref{eq_g2_second_order_2d_eq_Main}, we Taylor expand 
$
f^{\text{lab}}( \vec{x}, \vec{k}) $
about \( \vec{u}( \vec{x}) = 0 \). Since \( g( \vec{x}, \vec{k}) \) depends linearly on the perturbation, only the zeroth order term in the expansion contributes, allowing us to replace \( f^{\text{lab}} \) by
$
f^0 = \frac{1}{e^{\beta(\varepsilon_{\vec{k}} - \mu( \vec{x}))} + 1}$ in the above formulae.

\section{The Transport Coefficients}
In this section, we calculate the transport coefficients and demonstrate that they obey Onsager's reciprocity relations. The total currents can be written using the distribution function obtained in the previous section. Taking into account the phase factor $D_k$, we write the total currents as, \begin{subequations}
\label{eq_Currents and fluxes}
\begin{align}
j_i^N  &= -e \int [d\vec{k}] D_{\vec{k}} f( \vec{x}, \vec{k}) 
\mathit{\dot{x}}_i \label{eq_j_i} \\ 
j^Q_i &= \int [d\vec{k}] D_{\vec{k}} (\epsilon_k - \mu( \vec{x})) f( \vec{x}, \vec{k})
\mathit{\dot{x}}_i \label{eq_j_i^Q} \\ 
\pi_{i,j} &= \hbar\int [d\vec{k}] D_{\vec{k}} f( \vec{x}, \vec{k}) k_i 
\mathit{\dot{x}}_j, \label{eq_pi_ij}
\end{align}
\end{subequations} 
where $\vec{[dk]}=\frac{k_xk_y}{(2\pi)^2}$ and $D_{\vec{k}}=\frac{1}{1+\frac{e}{\hbar}\vec{B}\cdot \vec{\Omega_{\vec{k}}}}$ is the volume phase factor. The transport coefficients can be read off by substituting the distribution function value into the above equation. We assume in this section that $\tau_{ee}$ is dependent on $\vec{k}$ only through $\varepsilon_{\vec{k}}$. The transport coefficients to first order in $\tau_{ee}$ are 

\begin{subequations}
\begin{align}
L^{3}_{i,lj}&=e \int [d\vec{k}]\frac{ \tau_{ee} }{\hbar D_{\vec{k}}}\frac{\partial f^{0}}{\partial \varepsilon_{\vec{k}}} \pdv{\varepsilon_{\vec{k}}}{k_i}\pdv{\varepsilon_{\vec{k}}}{k_j} k_l
\label{eq_L_3_B_2d_first}
\\
L^{6}_{i,lj}&=- \int [d\vec{k}]\frac{ \tau_{ee} }{\hbar D_{\vec{k}}}(\varepsilon_{\vec{k}}-\mu)\frac{\partial f^{0}}{\partial \varepsilon_{\vec{k}}} \pdv{\varepsilon_{\vec{k}}}{k_i}\pdv{\varepsilon_{\vec{k}}}{k_j} k_l 
\label{eq_L_6_B_2d_first}
\\
L^{7}_{ij,l}&=e \int [d\vec{k}]\frac{ \tau_{ee} }{\hbar D_{\vec{k}} }\frac{\partial f^{0}}{\partial \varepsilon_{\vec{k}}} \pdv{\varepsilon_{\vec{k}}}{k_j}\pdv{\varepsilon_{\vec{k}}}{k_l} k_i
\label{eq_L_7_B_2d_first}
\\
L^{8}_{ij,l}&=- \int [d\vec{k}]\frac{ \tau_{ee} }{\hbar D_{\vec{k}} }\frac{\varepsilon_{\vec{k}}-\mu}{T
}\frac{\partial f^{0}}{\partial \varepsilon_{\vec{k}}} \pdv{\varepsilon_{\vec{k}}}{k_j}\pdv{\varepsilon_{\vec{k}}}{k_l} k_i.
\label{eq_L_8_B_2d_first}
\end{align}
\end{subequations}
We have suppressed the dependence of transport coefficients on $\vec{B}$, $\vec{\Omega_k}$, and $\varepsilon_{\vec{k}}$ in the LHS for brevity. On exchanging $i$ and $l$ in equations \ref{eq_L_7_B_2d_first} and \ref{eq_L_8_B_2d_first}, taking $\vec{\Omega_k}\rightarrow -\vec{\Omega_{-k}}$, $\varepsilon_{\vec{k}}\rightarrow \varepsilon_{-\vec{k}}$ and  $\vec{B}\rightarrow-\vec{B}$, and changing the variable of integration from $\vec{k}$ to $-\vec{k}$, it is readily seen that the Onsager's reciprocity relations in equations \ref{eq_Onsager_relatiosA} and \ref{eq_Onsager_relatiosB} are satisfied. In systems with inversion symmetry we have $\varepsilon_{\vec{k}}= \varepsilon_{\vec{-k}}$ and $\vec{\Omega_{k}}=\vec{\Omega_{-k}}$, making the integrand odd under $\vec{k}\rightarrow-\vec{k}$. Thus, the formulae obtained respect Curie's symmetry principle as the cross-tensor coefficients vanish in centrosymmetric systems.
We calculate the viscosity tensor in a similar manner to obtain, 

\begin{align}
L^{9}_{ij,ml}=- \int [d\vec{k}]\frac{ \tau_{ee} }{ D_{\vec{k}} }\frac{\partial f^{0}}{\partial \varepsilon_{\vec{k}}} \pdv{\varepsilon_{\vec{k}}}{k_j}\pdv{\varepsilon_{\vec{k}}}{k_l} k_ik_m
\label{eq_L_9_B_2d_first}
\end{align}  
The formula obeys the Onsager relation in equation \ref{eq_Onsager_relatiosC}. For isotropic Berry Curvature and dispersion, $\vec{\Omega_{k}}=\vec{\Omega}_{|\vec{k}|}$, $\varepsilon_{\vec{k}}=\varepsilon_{|\vec{k}|}$ and $\pdv{\varepsilon_{\vec{k}}}{k_m}=\pdv{\varepsilon_{k}}{k}\frac{k_m}{k}$the formula for $L^9$ simplifies to

\begin{align}
L^{9}_{ij,ml}=- \int [d\vec{k}]\frac{ \tau_{ee} }{ D_{k}}\frac{\partial f^{0}}{\partial \varepsilon_{k}} \pdv{\varepsilon_{k}}{k}\pdv{\varepsilon_{k}}{k} \frac{k_jk_l}{k^2}k_ik_m
\end{align}

This suggests that in the isotropic case, any component of the viscosity tensor with one index distinct and the other three equal vanishes implying that there is no odd viscosity component. It is noteworthy that in isotropic 2D systems with time-reversal breaking fields, the odd viscosity components can be non-zero\cite{Avron1998}. 

The second order dependence of transport coefficients on $\tau_{ee}$ is given by, 
\begin{widetext}
\begin{subequations}
 \begin{align}
  L^{3}_{i,lj}&=\frac{e^2}{\hbar^3}\int[d\vec{k}]\frac{\partial \varepsilon_{\vec{k}}}{\partial k_i}\left[
 \frac{\tau_{ee}^2}{D_{\vec{k}}} \epsilon_{pqr} \frac{\partial \varepsilon_{\vec{k}}}{\partial k_q}\pdv{\left(\frac{1}{D_{\vec{k}}}\right)}{k_p}\frac{\partial{f^{0} }}{{\partial\varepsilon_{\vec{k}} }}\pdv{\varepsilon_{\vec{k}}}{k_j}k_l  \right.+ \left. \frac{\tau_{ee}^2}{D_{\vec{k}}^2}\frac{\partial{f^{0} }}{{\partial\varepsilon_{\vec{k}}}}\epsilon_{pqr} \frac{\partial \varepsilon_{\vec{k}}}{\partial k_q}k_l\pdv{\varepsilon_{\vec{k}}}{k_j}{k_p} \right. 
 +\left.\frac{\tau_{ee}^2}{D_{\vec{k}}^2}\frac{\partial{f^{0} }}{{\partial\varepsilon_{\vec{k}} }}\epsilon_{lqr}\frac{\partial\varepsilon_{\vec{k}}}{\partial k_q}\frac{\partial \varepsilon_{\vec{k}}}{\partial k_j} \right]B^r \label{eq_L3_second_2d} 
 \end{align}
 \begin{align}
  L^{6}_{i,lj}&=-\frac{e}{\hbar^3}\int[d\vec{k}]\frac{\partial \varepsilon_{\vec{k}}}{\partial k_i}(\varepsilon_{\vec{k}}-\mu)\left[
 \frac{\tau_{ee}^2}{D_{\vec{k}}} \epsilon_{pqr} \frac{\partial \varepsilon_{\vec{k}}}{\partial k_q}\pdv{\left(\frac{1}{D_{\vec{k}}}\right)}{k_p}\frac{\partial{f^{0} }}{{\partial\varepsilon_{\vec{k}} }}\pdv{\varepsilon_{\vec{k}}}{k_j}k_l  \right. + \left. \frac{\tau_{ee}^2}{D_{\vec{k}}^2}\frac{\partial{f^{0} }}{{\partial\varepsilon_{\vec{k}}}}\epsilon_{pqr} \frac{\partial \varepsilon_{\vec{k}}}{\partial k_q}k_l\pdv{\varepsilon_{\vec{k}}}{k_j}{k_p} \right. 
 +\left.\frac{\tau_{ee}^2}{D_{\vec{k}}^2}\frac{\partial{f^{0} }}{{\partial\varepsilon_{\vec{k}} }}\epsilon_{lqr}\frac{\partial\varepsilon_{\vec{k}}}{\partial k_q}\frac{\partial \varepsilon_{\vec{k}}}{\partial k_j} \right]B^r
  \label{eq_L6_second_2d}
  \end{align}
  \begin{align}
  L^{7}_{ij,l}&=\frac{e^2}{\hbar^3}\int[d\vec{k}]k_i\frac{\partial \varepsilon_{\vec{k}}}{\partial k_j}\left[
 \frac{\tau_{ee}^2}{D_{\vec{k}}} \epsilon_{pqr} \frac{\partial \varepsilon_{\vec{k}}}{\partial k_q}\pdv{\left(\frac{1}{D_{\vec{k}}}\right)}{k_p}\frac{\partial{f^{0} }}{{\partial\varepsilon_{\vec{k}} }}\pdv{\varepsilon_{\vec{k}}}{k_l}   \right. + \left. \frac{\tau_{ee}^2}{D_{\vec{k}}^2}\frac{\partial{f^{0} }}{{\partial\varepsilon_{\vec{k}}}}\epsilon_{pqr} \frac{\partial \varepsilon_{\vec{k}}}{\partial k_q}\pdv{\varepsilon_{\vec{k}}}{k_l}{k_p} \right]B^r
 \label{eq_L7_second_2d}\end{align}
 \begin{align}
  L^{8}_{ij,l}&=-\frac{e}{\hbar^3}\int[d\vec{k}]k_i\frac{\partial \varepsilon_{\vec{k}}}{\partial k_j}\frac{\varepsilon_{\vec{k}}-\mu}{T}\left[
 \frac{\tau_{ee}^2}{D_{\vec{k}}} \epsilon_{pqr} \frac{\partial \varepsilon_{\vec{k}}}{\partial k_q}\pdv{\left(\frac{1}{D_{\vec{k}}}\right)}{k_p}\frac{\partial{f^{0} }}{{\partial\varepsilon_{\vec{k}} }}\pdv{\varepsilon_{\vec{k}}}{k_l}   \right. + \left. \frac{\tau_{ee}^2}{D_{\vec{k}}^2}\frac{\partial{f^{0} }}{{\partial\varepsilon_{\vec{k}}}}\epsilon_{pqr} \frac{\partial \varepsilon_{\vec{k}}}{\partial k_q}\pdv{\varepsilon_{\vec{k}}}{k_l}{k_p} \right]B^r
 \label{eq_L8_second_2d}
 \end{align}
\begin{align}
  L^{9}_{ij,lm}&=-\frac{e}{\hbar^2}\int[d\vec{k}]\frac{\partial \varepsilon_{\vec{k}}}{\partial k_j}k_i\left[
 \frac{\tau_{ee}^2}{D_{\vec{k}}} \epsilon_{pqr} \frac{\partial \varepsilon_{\vec{k}}}{\partial k_q}\pdv{\left(\frac{1}{D_{\vec{k}}}\right)}{k_p}\frac{\partial{f^{0} }}{{\partial\varepsilon_{\vec{k}} }}\pdv{\varepsilon_{\vec{k}}}{k_m}k_l  \right.+ \left. \frac{\tau_{ee}^2}{D_{\vec{k}}^2}\frac{\partial{f^{0} }}{{\partial\varepsilon_{\vec{k}}}}\epsilon_{pqr} \frac{\partial \varepsilon_{\vec{k}}}{\partial k_q}k_l\pdv{\varepsilon_{\vec{k}}}{k_m}{k_p} \right. 
 +\left.\frac{\tau_{ee}^2}{D_{\vec{k}}^2}\frac{\partial{f^{0}}}{{\partial\varepsilon_{\vec{k}} }}\epsilon_{lqr}\frac{\partial\varepsilon_{\vec{k}}}{\partial k_q}\frac{\partial \varepsilon_{\vec{k}}}{\partial k_m} \right]B^r
 \label{eq_L9_second_2d}
 \end{align}
 \end{subequations}
     
\end{widetext}
 It can be shown that the above formulae obey Onsager's relations (see Appendix \hyperref[Appendix_Onsager_second]{A4}). The cross-tensor coefficients can be seen to vanish when the system is centrosymmetric, as the integrand becomes odd under $\vec{k} \rightarrow -\vec{k}$, obeying Curie's principle.

\section{Conclusion}
In this paper, we have obtained an expression for the contribution of the Berry curvature to the viscosity tensor. The contribution arises due to the internal angular momentum of Bloch wavepackets arising from the Berry curvature, which renders the stress tensor asymmetric. As a result, the Berry curvature produces odd components of the viscosity tensor.  We have studied the effect of a magnetic field and Berry curvature on the vortical transport coefficients. Starting from microscopic reversibility, we have extended the Onsager relations for the vortical coefficients to include the Berry curvature and magnetic field. Working in the semiclassical approximation, we have shown that the contribution of the magnetic field to the vortical coefficients comes from the non-equilibrium part of the distribution function. We have obtained expressions for the distribution function and hence also the vortical coefficients to second order in the electron-electron scattering time and have shown explicitly that they are consistent with the Onsager relations we have derived. 

\section*{  ACKNOWLEDGMENTS}
We thank Nisarg Chadha, Archisman Panigrahi, 
and Sriram Ramaswamy for insightful discussions. A.S. acknowledges support from the Kishore Vaigyanik Protsahan Yojana from the Department of Science and 
Technology (DST), Government of India. S. M. thanks the Department of Science and 
Technology (DST), Government of India for support.

\onecolumngrid
\section*{Appendix}
\subsection*{A1. Deriving the Onsager's Relations} 
\label{Appendix_Onsager_der}
In this appendix, we derive Onsager's relations in the presence of a magnetic field for a system with non-vanishing Berry curvature. 
\subsubsection*{A1.1. State variables and their fluctuations}
The state variables are divided into two categories- $A_i$ and $B_j$(with $A_i$ being even under time-reversal and $B_j$ being odd under time reversal), with corresponding fluctuations from mean values being denoted by  $\alpha_i$ and $\beta_j$.  The set of time-reversal symmetry-breaking fields $\{\vec{H}\}$ transform under time reversal as: $ \{\vec{H}\}\rightarrow \Theta\left(\{\vec{H}\}\right)$. The change in entropy due to fluctuations in state variables can be written as:\cite{GrootBook2013,Groot1954,Chadha_Mukerjee2024}

\begin{align}\nonumber
\Delta S(\alpha,\beta )= -\frac{1}{2}\int &\int d \vec{x}d \vec{x'}\bigl(G_{ij}( \vec{x}, \vec{x'},\{\vec{H}\})\alpha_{i}( \vec{x})\alpha_{j}( \vec{x'})   + H_{pq}( \vec{x}, \vec{x'},\{\vec{H}\})\beta_{p}( \vec{x})\beta_{q}( \vec{x'}) \\ +  &U_{rs}( \vec{x}, \vec{x'},\{\vec{H}\})\alpha_{r}
( \vec{x})\beta_{s}( \vec{x'}) + V_{ab}( \vec{x}, \vec{x'},\{\vec{H}\})\beta_{a}( \vec{x})\alpha_{b}( \vec{x'}) \bigr), \end{align}with repeated indices being summed over. Non-negativity of $\Delta S$ requires that $G$ and $H$ be symmetric semi-positive definite matrices of appropriate dimensions. We have the freedom to impose $U^{T}=V$.  Since, $\Delta S$ is even under the reversal of velocities of particles and $ \{\vec{H}\}\xrightarrow{T.R.} \Theta\left(\{\vec{H}\}\right)$, we must have: 
\begin{subequations}
\begin{align}
G_{ij}( \vec{x}, \vec{x'},\Theta\left(\{\vec{H}\}\right)) =G_{ij}( \vec{x}, \vec{x'},\{\vec{H}\}) \\
H_{pq}( \vec{x}, \vec{x'},\Theta\left(\{\vec{H}\}\right)) =H_{pq}( \vec{x}, \vec{x'},\{\vec{H}\}) \\ 
U_{rs}( \vec{x}, \vec{x'},\Theta\left(\{\vec{H}\}\right)) =-U_{rs}( \vec{x}, \vec{x'},\{\vec{H}\})  \\ 
V_{ab}( \vec{x}, \vec{x'},\Theta\left(\{\vec{H}\}\right)) =-V_{ab}( \vec{x}, \vec{x'},\{\vec{H}\}), \end{align}
\end{subequations}
Using the Boltzmann entropy postulate, the probability of particular fluctuation configuration\cite{Gurzhi1968} :
\begin{equation}
    P\bigl(\{\alpha,\beta\} \bigr) \propto \exp(\Delta S(\alpha,\beta)  /k_B)
\end{equation}
The constant of proportionality can be found by appropriately normalizing the probability distribution,  $\int \Pi_{ij} D\bigl(\alpha_i,\beta_j \bigr)  P\bigl(\{\alpha_i,\beta_j\} \bigr)=1$(where $D$ denotes that we are performing a  functional integral)
\begin{equation}
    P\bigl(\{\alpha,\beta\} \bigr)= \frac{\exp(\Delta S(\alpha,\beta)/k_B)}{\int \Pi_{ij}D(\Tilde{\alpha_i},\Tilde{\beta_j})\exp(\Delta S(\Tilde{\alpha}_i,\Tilde{\beta}_j)/k_B) }
\end{equation}
The corresponding conjugate variables are given by:\\
\begin{subequations}
    \begin{align}
X_i( \vec{x})=\frac{\partial\Delta S(\alpha,\beta)}{\partial \alpha_i( \vec{x})}=k_B\frac{\partial\bigl( \exp({\Delta S/k_B})\bigr)}{\partial \alpha_i( \vec{x})}  \exp(-\Delta S(\alpha,\beta)/k_B) \\
Y_m( \vec{x})=\frac{\partial\Delta S(\alpha,\beta)}{\partial\beta_m( \vec{x})}=k_B\frac{\partial\bigl( \exp({\Delta S/k_B})\bigr)}{\partial \alpha_i( \vec{x})}  \exp(-\Delta S(\alpha,\beta)/k_B)
\end{align}
\end{subequations}
We next find various expectation values using the above definition and the Boltzmann entropy postulate.
\begin{equation}
    \bigl<\alpha_i( \vec{x}) X_j( \vec{x'})\bigr>= \frac{\int \Pi_{pq}D(\alpha_p,\beta_q)\alpha_i( \vec{x}) X_j( \vec{x'}) \exp(\Delta S(\alpha,\beta)/k_B)}{\int \Pi_{pq}D(\Tilde{\alpha_p},\Tilde{\beta_q})\exp(\Delta S(\Tilde{\alpha},\Tilde{\beta})/k_B)}
\end{equation}
Writing $\exp({\Delta S(\alpha,\beta)/k_B})=f(\alpha,\beta)$, we see:
\begin{equation}
    \frac{\bigl<\alpha_i( \vec{x}) X_j( \vec{x'})\bigr>}{k_B}= \frac{\int \Pi_{pq}D(\alpha_p,\beta_q)\alpha_i( \vec{x})\frac{\partial f(\alpha,\beta)}{\partial \alpha_j( \vec{x}')} }{\int \Pi_{pq}D(\Tilde{\alpha_p},\Tilde{\beta_q})f(\Tilde{\alpha},\Tilde{\beta})}
\end{equation}
Integrating by parts in the numerator on RHS and assuming $f\rightarrow0$ as $\alpha \rightarrow \pm\infty$, the expectation values simplifies to
\begin{equation}
    \bigl<\alpha_i( \vec{x}) X_j( \vec{x'})\bigr>=-k_{B}\delta_{ij} \delta( \vec{x}- \vec{x}') 
    \label{eq_alpha_X_expectation}
\end{equation}
Similarly,
\begin{eqnarray} 
    \bigl<\beta_i( \vec{x}) Y_j( \vec{x'})\bigr>=-k_{B}\delta_{ij} \delta( \vec{x}- \vec{x}') ;\ \ \
    \bigl<\alpha_i( \vec{x}) Y_j( \vec{x'})\bigr>=\bigl<\beta_i( \vec{x}) X_j( \vec{x'})\bigr>=0 
    \label{eq_beta_Y_expectation}
\end{eqnarray}
Next we define the derivative operator $\mathscr{D}{( \vec{x}')}=\sum_{u,v,w} a_{ijk} \delta'^u_i\delta'^v_j\delta'^w_k$. The subscript on $\delta$ denotes the spatial coordinates, and the superscript denotes the number of times the derivative is applied. Hitting on both sides of equations \ref{eq_alpha_X_expectation} and \ref{eq_beta_Y_expectation} with the derivative operator we get, 
\begin{subequations}\label{eq_various_exp}
\begin{align} \bigl<\alpha_i( \vec{x})\mathscr{D}( \vec{x}') X_j( \vec{x'})\bigr>&=-k_{B}\delta_{ij} \mathscr{D}( \vec{x}')\delta( \vec{x}- \vec{x}')  \label{eq_D_opra}\\
\bigl<\beta_i( \vec{x})\mathscr{D}( \vec{x}') Y_j( \vec{x'})\bigr>&=-k_{B}\delta_{ij} \mathscr{D}( \vec{x}')\delta( \vec{x}- \vec{x}') \label{eq_D_oprb}\\
\bigl<\alpha_i( \vec{x})\mathscr{D} ( \vec{x}')Y_j( \vec{x'})\bigr>&=0 \label{eq_D_oprc}\\
\bigl<\beta_i( \vec{x})\mathscr{D}( \vec{x}') X_j( \vec{x'})\bigr>&=0
\label{eq_D_oprd}
\end{align}
\end{subequations}
\subsubsection*{A1.2. Onsager's Reciprocity Relations in the presence of Berry Curvature}

Before deriving Onsager's reciprocity relations, we state without proof a few important equations coming reversibility of equations of motion, when the time reversal-breaking fields are appropriately transformed, \cite{Onsager1931,Groot1954,GrootBook2013}

\begin{subequations}
\begin{align}\label{eq_alpha_alpha_expectaion}
\bigl<\alpha_i( \vec{x},t,\{\vec{H}\})\frac{\partial }{\partial t}\bigl(\alpha_j( \vec{x}',t,\{\vec{H}\})\bigr)\bigr>=\bigl<\alpha_j( \vec{x}',t,\Theta\left(\{\vec{H}\}\right))\frac{\partial }{\partial t}\bigl(\alpha_i( \vec{x},t,\Theta\left(\{\vec{H}\}\right))\bigr)\bigr> \\ \label{eq_beta_beta_expectaion}
\bigl<\beta_i( \vec{x},t,\{\vec{H}\})\frac{\partial }{\partial t}\bigl(\beta_j( \vec{x}',t,\{\vec{H}\})\bigr)\bigr>=\bigl<\beta_j( \vec{x}',t,\Theta\left(\{\vec{H}\}\right))\frac{\partial }{\partial t}\bigl(\beta_i( \vec{x},t,\Theta\left(\{\vec{H}\}\right))\bigr)\bigr> \\ \label{eq_alpha_beta_expectaion}
\bigl<\alpha_i( \vec{x},t,\{\vec{H}\})\frac{\partial }{\partial t}\bigl(\beta_j( \vec{x}',t,\{\vec{H}\})\bigr)\bigr>=-\bigl<\beta_j( \vec{x}',t,\Theta\left(\{\vec{H}\}\right))\frac{\partial }{\partial t}\bigl(\alpha_i( \vec{x},t,\Theta\left(\{\vec{H}\}\right))\bigr)\bigr>
\end{align}
\end{subequations}
As described in the main text, $\vec{B}$ and $\boldsymbol{\Omega}_{\vec{k}}$ break time-reversal and should be transformed in the RHS of the above equations.

The steps outlined in the derivation are similar to those in Refs. \cite{Chadha_Mukerjee2024,Groot1954}. For our system, we have the number density of electrons($\bar{n}$), entropy density($\bar{s}$) as time-reversal even, and momentum density($\bar{\vec{g}}$) as time-reversal odd state variables. Corresponding currents are $\vec{j}^N$, $\vec{j}^Q$ and $\pi$.
We derive Onsager's relations for diagonal coefficients. In equation \ref{eq_alpha_alpha_expectaion}, we set $\alpha_i=\alpha_j=\Delta n$.
\begin{align}
\bigl<\Delta n( \vec{x},t,\{\vec{H}\})\frac{\partial }{\partial t}\bigl(\Delta n( \vec{x}',t,\{\vec{H}\})\bigr)\bigr>=\bigl<\Delta n( \vec{x}',t,\Theta(\{\vec{H}\}))\frac{\partial }{\partial t}\bigl(\Delta n( \vec{x},t,\Theta(\{\vec{H}\}))\bigr)\bigr> 
\end{align}
Using equations \ref{eq_continuity_charge}, 
\begin{align}
\bigl<\Delta n( \vec{x},\{\vec{H}\})\nabla' \cdot \vec{j^N}\bigr>&=\bigl<\Delta n'( \vec{x},\{\vec{H}\})\nabla \cdot \vec{J^N}\bigr>
\end{align}
Substituting the currents from equation \ref{eq_linear_charge} 
\begin{align}\nonumber&
\bigl<\Delta n( \vec{x},\Theta(\{\vec{H}\}))\partial_i'\bigl(-L^1_{i,j}( \vec{x}',\{\vec{H}\}) \partial'_j \phi - L^2_{i,j}( \vec{x}',\{\vec{H}\}) \partial'_j T -L^3_{i,jk}( \vec{x}',\{\vec{H}\}) \partial'_k u_j\bigr)\bigr> \\ =&\bigl<\Delta n( \vec{x}',\Theta(\{\vec{H}\}))\partial_i\bigl(-L^1_{i,j}( \vec{x},\Theta(\{\vec{H}\})) \partial_j \phi -  L^2_{i,j}( \vec{x},\Theta(\{\vec{H}\})) \partial_j T - L^3_{i,jk}( \vec{x},\Theta(\{\vec{H}\}))\partial_k u_j\bigr)\bigr> 
\end{align}
 Using \ref{eq_D_opra}, \ref{eq_D_oprb}, \ref{eq_D_oprc} and \ref{eq_D_oprd} we see, that only the expectation value of $\Delta n$ with position derivatives of $\phi$ would survive.
\begin{align}\nonumber
  &\bigl<\Delta n( \vec{x},\Theta(\{\vec{H}\}))L^1_{i,j}( \vec{x}',\{\vec{H}\})\partial_i' \partial'_j \phi\bigr>  =\bigl<\Delta n( \vec{x}',\Theta(\{\vec{H}\}))L^1_{i,j}( \vec{x},\Theta(\{\vec{H}\})) \partial_i\partial_j \phi \bigr>   
  \\
\implies&\partial_i' ( L^1_{i,j}( \vec{x}',\{\vec{H}\}) \partial_j'\delta( \vec{x}-  \vec{x}') )=  \partial_i \left( L^1_{i,j}( \vec{x},\Theta(\{\vec{H}\})) \partial_j \delta( \vec{x} -  \vec{x}') \right)
\end{align}
Multiplying by the arbitrary function $f( \vec{x}')$ on both sides and integrating with respect to $ \vec{x}'$\cite{Chadha_Mukerjee2024}.
\begin{equation}\int d \vec{x}'f( \vec{x}')\partial_i'\bigl(L^1_{i,j}( \vec{x}',\{\vec{H}\})
   \partial_j'\delta( \vec{x}-  \vec{x}')\bigr) = \int d \vec{x}' f( \vec{x}')\partial_i \bigl(L^1_{i,j}( \vec{x},\Theta(\{\vec{H}\})) \partial_j \delta( \vec{x} -  \vec{x}') \bigr)
\end{equation}
Integrating by parts(twice) on LHS and rearranging the integrand on the right.
\begin{equation}\int d \vec{x}'\delta( \vec{x}-  \vec{x}')\partial_j'   \bigl(L^1_{i,j}( \vec{x}',\{\vec{H}\})\partial_i'f( \vec{x}') \bigr)
= \int d \vec{x}' \partial_i \bigl(L^1_{i,j}( \vec{x},\Theta(\{\vec{H}\}))\partial_j \delta( \vec{x} -  \vec{x}') f( \vec{x}')\bigr)
\end{equation}
Integrating over delta functions on both sides.
\begin{equation} \partial_j   \bigl(L^1_{i,j}( \vec{x},\{\vec{H}\})\partial_i f( \vec{x}) \bigr)
    =\partial_i \bigl(L^1_{i,j}( \vec{x},\Theta(\{\vec{H}\})) \partial_j f( \vec{x}) \bigr)
\end{equation}
We expand the derivative using the product rule. Then we use the fact that $\partial_i\partial_j f( \vec{x})=\partial_j\partial_i f( \vec{x})$ and equate the coefficients of various spatial derivatives on LHS and RHS.
\begin{align}\nonumber& \partial_j   \bigl(L^1_{i,j}( \vec{x},\{\vec{H}\}) \bigr)\bigl(\partial_if( \vec{x}) \bigr)+ L^1_{i,j}( \vec{x},\{\vec{H}\})\partial_j   \partial_if( \vec{x})\\
=& \partial_i \bigl(L^1_{i,j}( \vec{x},\Theta(\{\vec{H}\})) \bigr)\partial_j f( \vec{x}) \bigr)+L^1_{i,j}( \vec{x},\Theta(\{\vec{H}\})) \partial_i  \partial_jf( \vec{x}) 
\end{align}
\begin{subequations}
\begin{equation}\implies\partial_j   \bigl(L^1_{i,j}( \vec{x},\{\vec{H}\}) \bigr)\bigl(\partial_if( \vec{x}) \bigr)
= \partial_i \bigl(L^1_{i,j}( \vec{x},\Theta(\{\vec{H}\})) \bigr)\partial_j f( \vec{x}) \bigr) \text{ and }
\label{eq_deriv_exp_a}
\end{equation}
\begin{equation}L^1_{i,j}( \vec{x},\{\vec{H}\})+L^1_{j,i}( \vec{x},\{\vec{H}\})
= L^1_{i,j}( \vec{x},\Theta(\{\vec{H}\}))+L^1_{j,i}( \vec{x},\Theta(\{\vec{H}\}))
\end{equation}
\end{subequations}
Exchanging dummy indices $i$ and $j$ in LHS of \ref{eq_deriv_exp_a}.
\begin{equation} \partial_i   \bigl(L^1_{j,i}( \vec{x},\{\vec{H}\})\partial_j f( \vec{x}) \bigr)
    =\partial_i \bigl(L^1_{i,j}( \vec{x},\Theta(\{\vec{H}\})) \partial_j f( \vec{x}) \bigr)
\end{equation}
Choosing $f( \vec{x})=x_k$, we see ,
\begin{equation} \partial_i   \bigl(L^1_{k,i}( \vec{x},\{\vec{H}\}) \bigr)
    =\partial_i \bigl(L^1_{i,k}( \vec{x},\Theta(\{\vec{H}\}))  \bigr)
\end{equation}
Fixing $k$ and considering $i$ as indices of a vector,  we see that $L^1_{i,k}( \vec{x},\{\vec{H}\})$ and $L^1_{k,i}( \vec{x},\Theta(\{\vec{H}\}))$ have equal divergences and, therefore, must differ by curl of some vector. However, we know outside the sample both $L^1_{i,k}$ and $L^1_{k,i}$ equal zero. Therefore we must have,
\begin{equation}
 L^1_{i,j}( \vec{x},\{\vec{H}\})=L^1_{j,i}( \vec{x},\Theta(\{\vec{H}\}))   
\end{equation}
In the case of interest, 
\begin{equation}
 L^1_{i,j}( \vec{x},\vec{B};\{\vec{\Omega_k}\}, \{\varepsilon_{\vec{k}}\} )=L^1_{j,i}( \vec{x},\vec{B};\{-\vec{\Omega_{-k}}\}, \{\varepsilon_{-\vec{k}}\} ) 
 \label{eq_onsagerL1}
\end{equation}
 Similarly, we can prove Onsager's relations for other diagonal coefficients. By setting  $\alpha_i=\alpha_j=\Delta s $ in equation \ref{eq_alpha_alpha_expectaion} and   $\beta_i=\Delta g_i$  and $\beta_j=\Delta g_j$ in equation \ref{eq_beta_beta_expectaion} we get the following relations.
\begin{empheq}{align}
 L^5_{i,j}( \vec{x},\vec{B};\{\vec{\Omega_k}\}, \{\varepsilon_{\vec{k}}\} )&=L^5_{j,i}( \vec{x},\vec{B};\{-\vec{\Omega_{-k}}\}, \{\varepsilon_{-\vec{k}}\} ) \label{eq_onsagerL4}  \\
 L^9_{il,jk}( \vec{x},\vec{B};\{\vec{\Omega_k}\}, \{\varepsilon_{\vec{k}}\} )&=L^9_{jk ,il}( \vec{x},\vec{B};\{-\vec{\Omega_{-k}}\}, \{\varepsilon_{-\vec{k}}\} )  
 \label{eq_onsagerL9}
\end{empheq}
A similar application of equations \ref{Transport_eq},  \ref{eq_various_exp} and \ref{eq_alpha_beta_expectaion} would give Onsager's reciprocity relations for off-diagonal coefficients.
\subsection*{A2. Simplification of RHS of equation \ref{eq_Boltzmann_simplified1}}\label{Appendix_RHS_Boltz}\begin{align}\nonumber
&\pdv{f^{\text{lab}}( \vec{x},\vec{k})}{ \vec{x}} = \nabla_{ \vec{x}} \mu{( \vec{x})}\frac{\partial{f^{\text{lab}}( \vec{x},\vec{k})}}{{\partial\mu({ \vec{x}}) }}+\nabla_{ \vec{x}} \beta{( \vec{x})}\frac{\partial{f^{\text{lab}}( \vec{x},\vec{k})}}{{\partial\beta({ \vec{x}}) }}+\nabla_{ \vec{x}} (\vec{u}({ \vec{x}})\cdot \vec{k})))\frac{\partial{f^{\text{lab}}( \vec{x},\vec{k})}}{{\partial(\vec{u}\cdot \vec{k}) }}
 \\ \nonumber =& -\nabla_{ \vec{x}} \mu{( \vec{x})}\frac{\partial{f^{\text{lab}}( \vec{x},\vec{k})}}{{\partial\varepsilon_{\vec{k}} }}-\frac{\nabla_{ \vec{x}} T{( \vec{x})}}{T( \vec{x})^2}\frac{\partial{f^{\text{lab}}( \vec{x},\vec{k})}}{{\partial\bigl(\beta({ \vec{x}})(\varepsilon_{\vec{k}}-\mu( \vec{x})-\hbar\vec{u}( \vec{x})\cdot \vec{k})\bigr) }}\bigl(\varepsilon_{\vec{k}}-\mu( \vec{x})-\hbar\vec{u}( \vec{x})\cdot \vec{k}\bigr) -\bigl(\hbar\nabla_{ \vec{x}} \vec{u}({ \vec{x}})\bigr)\cdot \vec{k}\frac{\partial{f^{\text{lab}}( \vec{x},\vec{k})}}{{\partial\varepsilon_{\vec{k}} }}
 \\ \nonumber  =& -\nabla_{ \vec{x}} \mu{( \vec{x})}\frac{\partial{f^{\text{lab}}( \vec{x},\vec{k})}}{{\partial\varepsilon_{\vec{k}} }}-\frac{\nabla_{ \vec{x}} T{( \vec{x})}}{T( \vec{x})}\frac{\partial{f^{\text{lab}}( \vec{x},\vec{k})}}{{\partial \varepsilon_{\vec{k}} }}\bigl(\varepsilon_{\vec{k}}-\mu( \vec{x}))-\bigl(\hbar\nabla_{ \vec{x}} \vec{u}({ \vec{x}})\bigr)\cdot \vec{k}\frac{\partial{f^{\text{lab}}( \vec{x},\vec{k})}}{{\partial\varepsilon_{\vec{k}} }}
 \\=& \frac{\partial{f^{\text{lab}}( \vec{x},\vec{k})}}{{\partial\varepsilon_{\vec{k}} }}\left(-\nabla_{ \vec{x}} \mu{( \vec{x})}-\frac{\nabla_{ \vec{x}} T{( \vec{x})}}{T( \vec{x})}\bigl(\varepsilon_{\vec{k}}-\mu( \vec{x}))-\bigl( \hbar\nabla_{ \vec{x}} \vec{u}({ \vec{x}}) \bigr)\vec{k} \right)
 \label{eq_der_r_f}
\end{align}
We assume that $\vec{u}( \vec{x})$ is linear(or sub-linear) in applied fields. In the third equality above, we have dropped the product of $\vec{u}( \vec{x})\cdot\vec{k} $ with $\nabla_{ \vec{x}}T( \vec{x})$ because we keep terms only up to first order in perturbing fields.
\begin{align}\nonumber
-\dot{ \vec{x}}\cdot \nabla_{ \vec{x}}f^{\text{lab}}( \vec{x},\vec{k})&=-\frac{\nabla_{\vec{k}}\varepsilon_{\vec{k}} +e \vec{E} \times \vec{\Omega}_k }{\hbar D_{\vec{k}}}\frac{\partial{f^{\text{lab}}( \vec{x},\vec{k})}}{{\partial\varepsilon_{\vec{k}} }}\left(-\nabla_{ \vec{x}} \mu{( \vec{x})}-\frac{\nabla_{ \vec{x}} T{( \vec{x})}}{T( \vec{x})}\bigl(\varepsilon_{\vec{k}}-\mu( \vec{x}))-\hbar\bigl(\nabla_{ \vec{x}} \vec{u}({ \vec{x}}) \bigr)\vec{k} \right) \\& =
\frac{1}{ \hbar D_{\vec{k}}} \frac{\partial{f^{\text{lab}}( \vec{x},\vec{k})}}{{\partial\varepsilon_{\vec{k}} }}\nabla_{\vec{k}}\varepsilon_{\vec{k}}\cdot\left(\nabla_{ \vec{x}} \mu{( \vec{x})}+\frac{\nabla_{ \vec{x}} T{( \vec{x})}}{T( \vec{x})}\bigl(\varepsilon_{\vec{k}}-\mu( \vec{x}))+\hbar\bigl(\nabla_{ \vec{x}} \vec{u}({ \vec{x}}) \bigr)\vec{k} \right) 
\end{align}
Again, we have kept terms only up to first order in perturbing fields. 
\begin{align}
    \nonumber -  \vec{\dot{k}} \cdot \nabla_{\vec{k}} f^{\text{lab}}( \vec{x},\vec{k}) &= \frac{e\vec{E} +\frac{e}{\hbar}\nabla_{\vec{k}} \varepsilon_{\vec{k}} \times \vec{B}}{\hbar D_{k} }\cdot\left(\nabla_{\vec{k}} \varepsilon_{\vec{k}}\frac{\partial{f^{\text{lab}}( \vec{x},\vec{k})}}{{ \partial\varepsilon_{\vec{k}}}}+\nabla_{\vec{k}}( \vec{u}({ \vec{x}})\cdot \vec{k})\frac{\partial{f^{\text{lab}}( \vec{x},\vec{k})}}{{\partial(\vec{u}\cdot \vec{k}) }}\right) \\ \nonumber  &= \frac{e\vec{E} +\frac{e}{\hbar}\nabla_{\vec{k}} \varepsilon_{\vec{k}} \times \vec{B}}{ \hbar D_{k} }\cdot\left(\nabla_{\vec{k}} \varepsilon_{\vec{k}}\frac{\partial{f^{\text{lab}}( \vec{x},\vec{k})}}{{ \partial\varepsilon_{\vec{k}}}}- \hbar\vec{u}({ \vec{x}})\frac{\partial{f^{\text{lab}}( \vec{x},\vec{k})}}{{\partial\varepsilon_{\vec{k}} }}\right) \\ \nonumber  &=  \frac{\partial{f^{\text{lab}}( \vec{x},\vec{k})}}{{\partial\varepsilon_{\vec{k}} }}\frac{e}{\hbar D_{k} }\left(\vec{E}\cdot\nabla_{\vec{k}} \varepsilon_{\vec{k}}-\left(\nabla_{\vec{k}} \varepsilon_{\vec{k}} \times \vec{B} \right)\cdot \vec{u}({ \vec{x}})\right)
     \nonumber  \\ &=  \frac{\partial{f^{\text{lab}}( \vec{x},\vec{k})}}{{\partial\varepsilon_{\vec{k}} }}\frac{ \nabla_{\vec{k}} \varepsilon_{\vec{k}}}{\hbar D_{k} }\cdot \bigl(e\vec{E}+e\vec{u}({ \vec{x}})\times \vec{B} \bigr)
\end{align}
Thus, equation RHS of \ref{eq_Boltzmann_simplified1} simplifies to 
\begin{align} 
 \frac{\partial{f^{\text{lab}}( \vec{x},\vec{k})}}{{\partial\varepsilon_{\vec{k}} }}\frac{\nabla_{\vec{k}} \varepsilon_{\vec{k}}}{\hbar D_{k} }\cdot \biggl(e\vec{E}+e \vec{u}({ \vec{x}})\times& \vec{B} +\nabla_{ \vec{x}} \mu{( \vec{x})}+\frac{\nabla_{ \vec{x}} T{( \vec{x})}}{T( \vec{x})}\bigl(\varepsilon_{\vec{k}}-\mu( \vec{x}))+\hbar\bigl(\nabla_{ \vec{x}} \vec{u}({ \vec{x}}) \bigr)\vec{k} \biggr)
\end{align}
\subsection*{A3. Calculating the Out-of-Equilibrium distribution function}\label{Appendix_out_of_eq}
Having simplified equation \ref{eq_Boltzmann_simplified1}, we proceed to solve it. Substituting for  $\dot{\vec{k}}$ and $\dot{ \vec{x}}$, the LHS of equation \ref{eq_Boltzmann_simplified2} simplifies to,
\begin{equation}
   \frac{g( \vec{x},\vec{k})}{\tau_{ee}}+\frac{
\frac{1}{\hbar} \vec{\nabla_k}\varepsilon_{\vec{k}} 
+ \frac{e}{\hbar} (\vec{E} \times \vec{\Omega_k}) 
}{ 1 + \frac{e}{\hbar} \vec{B} \cdot \vec{\Omega_k} }\cdot\pdv{g( \vec{x},\vec{k})}{ \vec{x}}-\frac{\frac{e}{\hbar} \vec{E} 
+ \frac{e}{\hbar^2} \vec{\nabla_k}\varepsilon_{\vec{k}} \times \vec{B} 
}{ 1 + \frac{e}{\hbar} \vec{B} \cdot \vec{\Omega_k} }\cdot\pdv{g( \vec{x},\vec{k})}{\vec{k}}
\end{equation}
Realizing that $g$ is $O(1)$ in the perturbing fields, we retain only the terms that are first order in perturbing fields.
\begin{equation}
   \frac{g( \vec{x},\vec{k})}{\tau_{ee}}+\frac{
\frac{1}{\hbar} \vec{\nabla_k}\varepsilon_{\vec{k}} 
}{ 1 + \frac{e}{\hbar} \vec{B} \cdot \vec{\Omega_k} }\cdot\pdv{g( \vec{x},\vec{k})}{ \vec{x}}-\frac{ \frac{e}{\hbar^2} \vec{\nabla_k}\varepsilon_{\vec{k}} \times \vec{B} 
}{ 1 + \frac{e}{\hbar} \vec{B} \cdot \vec{\Omega_k} }\cdot\pdv{g( \vec{x},\vec{k})}{\vec{k}}
\label{expr_LHS_simpified_Boltzmann_subst}
\end{equation}
\subsubsection*{A3.1 First order dependence}
The third term in the expression \ref{expr_LHS_simpified_Boltzmann_subst} can be estimated as
\begin{equation}
e\left(\nabla_{\vec{k}} \varepsilon_{\vec{k}} \times \vec{B}\right) \cdot \frac{\partial g( \vec{x}, \vec{k})}{\partial \vec{k}} 
\sim \left(\vec{v} \times e\vec{B}\right) \cdot \frac{\partial g( \vec{x}, \vec{k})}{\partial (m\vec{v})} 
\sim \frac{e \vec{B}}{m} g
\end{equation}
Thus, if $\omega \ll \frac{1}{\tau_{ee}}$, we can regard the third term as a perturbation~\cite{Panigrahi_Mukerjee2023}. The second term will be shown to be negligible self-consistently.
The equation \ref{eq_Boltzmann_simplified2} simplifies to,
\begin{equation}
   \frac{g( \vec{x},\vec{k})}{\tau_{ee}}+\frac{
\frac{1}{\hbar} \vec{\nabla_k}\varepsilon_{\vec{k}} 
}{ 1 + \frac{e}{\hbar} \vec{B} \cdot \vec{\Omega_k} }\cdot\pdv{g( \vec{x},\vec{k})}{ \vec{x}}-\frac{ \frac{e}{\hbar^2} \vec{\nabla_k}\varepsilon_{\vec{k}} \times \vec{B} 
}{ 1 + \frac{e}{\hbar} \vec{B} \cdot \vec{\Omega_k} }\cdot\pdv{g( \vec{x},\vec{k})}{\vec{k}}=\frac{1}{\hbar}\frac{\partial{f^{\text{lab}}( \vec{x},\vec{k})}}{{\partial\varepsilon_{\vec{k}} }}\frac{\nabla_{\vec{k}} \varepsilon_{\vec{k}}}{D_{k} }\cdot \vec{S}
\label{eq_Boltzmann_simplified3_App}
\end{equation}
Following the condition of a weak magnetic field, from equation \ref{eq_Boltzmann_simplified2} the first order term is immediately seen to be,
\begin{equation}
g_1( \vec{x}, \vec{k}) =\frac{ \tau_{ee} }{\hbar}\frac{\partial f^{\text{lab}}( \vec{x}, \vec{k})}{\partial \varepsilon_{\vec{k}}} \frac{\nabla_{\vec{k}} \varepsilon_{\vec{k}}}{D_{\vec{k}}} \cdot \vec{S},
\label{eq_g1_first_order_2d_App}
\end{equation}
where
 \begin{equation}
\vec{S}=e\vec{E}+e \vec{u}({ \vec{x}})\times \vec{B} +\nabla_{ \vec{x}} \mu{( \vec{x})} +\frac{\nabla_{ \vec{x}} T{( \vec{x})}}{T( \vec{x})}\bigl(\varepsilon_{\vec{k}}-\mu( \vec{x}))+\hbar \bigl(\nabla_{ \vec{x}} \vec{u}({ \vec{x}}) \bigr)\vec{k}
\end{equation}
  Since a Lorentz transformation at non-relativistic speeds can generate the term $\vec{u}\times\vec{B}$, it is treated as an external perturbing field whose spatial derivative is negligible. The fields $\mu( \vec{x})$, $T( \vec{x})$, and $\vec{u}( \vec{x})$ are assumed to be slowly varying, whose second spatial derivative can be dropped. Thus, we have self-consistently shown that we can ignore the second term in the expression \ref{expr_LHS_simpified_Boltzmann_subst}
up to linear order in the perturbing fields.
\subsubsection*{A3.2 Second order dependence}
In the simplified Boltzmann-transport equation \ref{eq_Boltzmann_simplified3_App}, we back-substitute the first-order solution  to obtain the second-order dependence \cite{Panigrahi_Mukerjee2023},
\begin{equation}
    \frac{g_2( \vec{x},\vec{k})}{\tau_{ee}}=\frac{e}{\hbar ^2D_{\vec{k}}}\left(\nabla_{\vec{k}}\varepsilon_{\vec{k}}\times \vec{B}\right)\cdot\frac{\partial g_1( \vec{x},\vec{k})}{\partial \vec{k}}
    \label{eq_g2_second_order_2d_eq_App}
\end{equation}

\paragraph{\texorpdfstring{When $\overrightarrow{S}$ is $\overrightarrow{k}$ independent,}{}}
As an example take $\vec{S}=e\vec{E}$.
\begin{equation}
\frac{\partial g_{1}( \vec{x},\vec{k}) }{\partial \vec{k}}=\frac{\partial\left(\frac{\tau_{ee}}{\hbar D_{\vec{k}}}\frac{\partial{f^{\text{lab}}( \vec{x},\vec{k})}}{{\partial\varepsilon_{\vec{k}} }}\nabla_{\vec{k}} \varepsilon_{\vec{k}}\cdot e\vec{E}\right)}{\partial \vec{k}}  \end{equation}
Now we use the vector calculus identity \begin{align}\nonumber\nabla (\alpha \vec{X} \cdot \vec{Y})&=\nabla(\alpha) (\vec{X}\cdot \vec{Y} )+\alpha\nabla (\vec{X}\cdot \vec{Y} ) \\ &=\nabla(\alpha) (\vec{X}\cdot \vec{Y} ) +\alpha (\vec{X}\cdot\nabla)\vec{Y}+ \alpha \vec{X}\times(\nabla\times\vec{Y})  +\alpha (\vec{Y}\cdot\nabla)\vec{X}+ \alpha \vec{Y}\times(\nabla\times\vec{X}) \label{eq_vec_identity}\end{align}
Taking $\alpha=\frac{\tau_{ee}}{\hbar D_{\vec{k}}}\frac{\partial{f^{\text{lab}}( \vec{x},\vec{k})}}{{\partial\varepsilon_{\vec{k}} }}\nabla_{\vec{k}}$,  $\vec{X}= \nabla_{\vec{k}} \varepsilon_{\vec{k}}$ and $\vec{Y}=e\vec{E}$ and noting that $\vec{Y}$ is $\vec{k}$ independent.
\begin{align} \nonumber
\frac{\partial g_1( \vec{x},\vec{k}) }{\partial \vec{k}}=&\frac{\partial \left(\frac{\tau_{ee}}{\hbar D_{\vec{k}}}\frac{\partial{f^{\text{lab}}( \vec{x},\vec{k})}}{{\partial\varepsilon_{\vec{k}} }}\right) }{\partial \vec{k}}(\nabla_{\vec{k}}\varepsilon_{\vec{k}}\cdot e\vec{E})+\left(\frac{\tau_{ee}}{\hbar D_{\vec{k}}}\frac{\partial{f^{\text{lab}}( \vec{x},\vec{k})}}{{\partial\varepsilon_{\vec{k}} }}\right)(e\vec{E}\cdot\nabla_{\vec{k}} )(\nabla_{\vec{k}}\varepsilon_{\vec{k}}) \\& +\nonumber\left(\frac{\tau_{ee}}{\hbar D_{\vec{k}}}\frac{\partial{f^{\text{lab}}( \vec{x},\vec{k})}}{{\partial\varepsilon_{\vec{k}} }}\right)e\vec{E}\times (\nabla_{\vec{k}}\times (\nabla_{\vec{k}}\varepsilon_{\vec{k}})) \\ =&\frac{\partial \left(\frac{\tau_{ee}}{\hbar D_{\vec{k}}}\frac{\partial{f^{\text{lab}}( \vec{x},\vec{k})}}{{\partial\varepsilon_{\vec{k}} }}\right) }{\partial \vec{k}}(\nabla_{\vec{k}}\varepsilon_{\vec{k}}\cdot e\vec{E})+\left(\frac{\tau_{ee}}{\hbar D_{\vec{k}}}\frac{\partial{f^{\text{lab}}( \vec{x},\vec{k})}}{{\partial\varepsilon_{\vec{k}} }}\right)(e\vec{E}\cdot\nabla_{\vec{k}} )(\nabla_{\vec{k}}\varepsilon_{\vec{k}}) 
\label{eq_delg1_del_k_S_E}
\end{align}
 The simplification for $\vec{S}=e\vec{u}\times\vec{B}$ and $\vec{S}=\nabla_{ \vec{x}}\mu( \vec{x})$ follows immediately as these too are $\vec{k}$ independent.
The full expression for $g_2( \vec{x},\vec{k})$ when $\vec{S}'=e\vec{E}+e\vec{u}\times\vec{B}+\nabla_{ \vec{x}}\mu( \vec{x})$ follows from equations \ref{eq_g2_second_order_2d_eq_App} and \ref{eq_delg1_del_k_S_E}, \begin{align}  \nonumber
 g_2( \vec{x},\vec{k}) =\frac{e\tau_{ee}}{\hbar^3D_{\vec{k}}}&\left(\left(\nabla_{\vec{k}}\varepsilon_{\vec{k}}\times \vec{B}\right)\cdot\frac{\partial \left(\frac{\tau_{ee}}{D_{\vec{k}}}\frac{\partial{f^{\text{lab}}( \vec{x},\vec{k})}}{{\partial\varepsilon_{\vec{k}} }}\right) }{\partial \vec{k}}\right)(\nabla_{\vec{k}}\varepsilon_{\vec{k}}\cdot \vec{S}')+\\&  \frac{e\tau_{ee}}{\hbar^3 D_{\vec{k}}}\left(\frac{\tau_{ee}}{D_{\vec{k}}}\frac{\partial{f^{\text{lab}}( \vec{x},\vec{k})}}{{\partial\varepsilon_{\vec{k}} }}\right)\left(\left(\nabla_{\vec{k}}\varepsilon_{\vec{k}}\times \vec{B}\right)\cdot((\vec{S}'\cdot\nabla_{\vec{k}} )\nabla_{\vec{k}}\varepsilon_{\vec{k}})\right)\label{eq_g_2_S_E}\end{align}

\paragraph{\texorpdfstring{When $\overrightarrow{S} = \dfrac{\overrightarrow{\nabla}_{ \vec{x}} T( \vec{x})}{T( \vec{x})} \left( \varepsilon_{\vec{k}} - \mu( \vec{x}) \right)$,}{}}We consider $\vec{S}=\frac{\nabla_{ \vec{x}} T{( \vec{x})}}{T( \vec{x})}\bigl(\varepsilon_{\vec{k}}-\mu( \vec{x}))$
\begin{equation}
\frac{\partial g_1( \vec{x},\vec{k}) }{\partial \vec{k}}=\frac{\partial\left(\frac{\tau_{ee}}{\hbar D_{\vec{k}}}\frac{\partial{f^{\text{lab}}( \vec{x},\vec{k})}}{{\partial\varepsilon_{\vec{k}} }}\nabla_{\vec{k}} \varepsilon_{\vec{k}}\cdot\nabla_{ \vec{x}} T{( \vec{x})}\frac{\bigl(\varepsilon_{\vec{k}}-\mu( \vec{x})\bigr)}{T( \vec{x})}\right)}{\partial \vec{k}} = \frac{\partial\left(\frac{\tau_{ee}}{\hbar D_{\vec{k}}}\frac{\partial{f^{\text{lab}}( \vec{x},\vec{k})}}{{\partial\varepsilon_{\vec{k}} }}\frac{\bigl(\varepsilon_{\vec{k}}-\mu( \vec{x})\bigr)}{T( \vec{x})}\nabla_{\vec{k}} \varepsilon_{\vec{k}}\cdot\nabla_{ \vec{x}} T{( \vec{x})}\right)}{\partial \vec{k}}  \end{equation}
We again use identity \ref{eq_vec_identity}. Taking $\alpha=\frac{\tau_{ee}}{\hbar D_{\vec{k}}}\frac{\partial{f^{\text{lab}}( \vec{x},\vec{k})}}{{\partial\varepsilon_{\vec{k}} }}\frac{\bigl(\varepsilon_{\vec{k}}-\mu( \vec{x})\bigr)}{T( \vec{x})}$,  $\vec{X}= \nabla_{\vec{k}} \varepsilon_{\vec{k}}$ and $\vec{Y}=\nabla_{ \vec{x}}T( \vec{x})$ and noting that that $\vec{Y}$ is independent of $\vec{k}$.
\begin{align} \nonumber
\frac{\partial g_1( \vec{x},\vec{k}) }{\partial \vec{k}}=\frac{\partial \left(\frac{\tau_{ee}}{\hbar D_{\vec{k}}}\frac{\partial{f^{\text{lab}}( \vec{x},\vec{k})}}{{\partial\varepsilon_{\vec{k}} }}\frac{\bigl(\varepsilon_{\vec{k}}-\mu( \vec{x})\bigr)}{T( \vec{x})}\right) }{\partial \vec{k}}&(\nabla_{\vec{k}}\varepsilon_{\vec{k}}\cdot \nabla_{ \vec{x}}T( \vec{x}))+\\ \nonumber &\left(\frac{\tau_{ee}}{\hbar D_{\vec{k}}}\frac{\partial{f^{\text{lab}}( \vec{x},\vec{k})}}{{\partial\varepsilon_{\vec{k}} }}\frac{\bigl(\varepsilon_{\vec{k}}-\mu( \vec{x})\bigr)}{T( \vec{x})}\right)(\nabla_{ \vec{x}}T( \vec{x})\cdot\nabla_{\vec{k}} )(\nabla_{\vec{k}}\varepsilon_{\vec{k}}) 
\end{align}
Expanding the $\frac{\partial}{\partial \vec{k}}$ on first term of RHS,
\begin{align} \nonumber
\frac{\partial g_1( \vec{x},\vec{k}) }{\partial \vec{k}}=\frac{\partial \left(\frac{\tau_{ee}}{\hbar D_{\vec{k}}}\frac{\partial{f^{\text{lab}}( \vec{x},\vec{k})}}{{\partial\varepsilon_{\vec{k}} }}\right) }{\partial \vec{k}}\frac{\bigl(\varepsilon_{\vec{k}}-\mu( \vec{x})\bigr)}{T( \vec{x})}(\nabla_{\vec{k}}\varepsilon_{\vec{k}}&\cdot \nabla_{ \vec{x}}T( \vec{x}))+\frac{\tau_{ee}}{\hbar D_{\vec{k}} T( \vec{x})}\frac{\partial{f^{\text{lab}}( \vec{x},\vec{k})}}{{\partial\varepsilon_{\vec{k}} }}\frac{\partial \varepsilon_{\vec{k}} }{\partial \vec{k}}(\nabla_{\vec{k}}\varepsilon_{\vec{k}}\cdot \nabla_{ \vec{x}}T( \vec{x}))\\ &+\left(\frac{\tau_{ee}}{\hbar D_{\vec{k}}}\frac{\partial{f^{\text{lab}}( \vec{x},\vec{k})}}{{\partial\varepsilon_{\vec{k}} }}\frac{\bigl(\varepsilon_{\vec{k}}-\mu( \vec{x})\bigr)}{T( \vec{x})}\right)(\nabla_{ \vec{x}}T( \vec{x})\cdot\nabla_{\vec{k}} )(\nabla_{\vec{k}}\varepsilon_{\vec{k}}) 
\end{align}
The second term in RHS of the above equation is parallel to $\nabla_{\vec{k}}\varepsilon_{\vec{k}}$ and hence perpendicular to $\nabla_{\vec{k}}\varepsilon_{\vec{k}}\times \vec{B}$. Therefore, it would vanish when substituted in equation \ref{eq_g2_second_order_2d_eq_App}. The full expression for $g_2( \vec{x},\vec{k})$ when $\vec{S}''=e\vec{E}+e\vec{u}\times\vec{B}+\nabla_{ \vec{x}}\mu( \vec{x})+\frac{\nabla_{ \vec{x}} T{( \vec{x})}}{T( \vec{x})}\bigl(\varepsilon_{\vec{k}}-\mu( \vec{x}))$ follows from \ref{eq_g2_second_order_2d_eq_App} and then adding the contribution from equation \ref{eq_g_2_S_E}\begin{align} \nonumber 
 g_2( \vec{x},\vec{k}) =\frac{e\tau_{ee}}{ \hbar^3 D_{\vec{k}}}&\left(\left(\nabla_{\vec{k}}\varepsilon_{\vec{k}}\times \vec{B}\right)\cdot\frac{\partial \left(\frac{\tau_{ee}}{ D_{\vec{k}}}\frac{\partial{f^{\text{lab}}( \vec{x},\vec{k})}}{{\partial\varepsilon_{\vec{k}} }}\right) }{\partial \vec{k}}\right)(\nabla_{\vec{k}}\varepsilon_{\vec{k}}\cdot \vec{S}'') \\ &+\frac{e\tau_{ee}}{\hbar^3 D_{\vec{k}}}\left(\frac{\tau_{ee}}{D_{\vec{k}}}\frac{\partial{f^{\text{lab}}( \vec{x},\vec{k})}}{{\partial\varepsilon_{\vec{k}} }}\right)\left(\left(\nabla_{\vec{k}}\varepsilon_{\vec{k}}\times \vec{B}\right)\cdot((\vec{S}''\cdot\nabla_{\vec{k}} )\nabla_{\vec{k}}\varepsilon_{\vec{k}})\right)\label{eq_g_2_S_E_mu}\end{align}

\paragraph{\texorpdfstring{When $\overrightarrow{S}=\hbar\bigl(\nabla_{ \vec{x}} \overrightarrow{u}({ \vec{x}}) \bigr)\cdot \overrightarrow{k}$}{}}
    Finally we take $\vec{S}=\hbar\bigl(\nabla_{ \vec{x}} \vec{u}({ \vec{x}}) \bigr)\vec{k}$.
 \begin{equation}
\frac{\partial g_{1}( \vec{x},\vec{k}) }{\partial \vec{k}}=\frac{\partial\left(\frac{\tau_{ee}}{\hbar D_{\vec{k}}}\frac{\partial{f^{\text{lab}}( \vec{x},\vec{k})}}{{\partial\varepsilon_{\vec{k}} }}\left(\hbar \nabla_{\vec{k}} \varepsilon_{\vec{k}}\cdot \bigl(\bigl(\nabla_{ \vec{x}} \vec{u}({ \vec{x}}) \bigr)\vec{k}\bigr)\right)\right)}{\partial \vec{k}}   \end{equation}
 We shall use the following vector identity: \begin{align}\nonumber\nabla (\alpha \vec{X} \cdot \vec{Y})&=\nabla(\alpha) (\vec{X}\cdot \vec{Y} )+\alpha\nabla (\vec{X}\cdot \vec{Y} )\\\nonumber &=\nabla(\alpha) (\vec{X}\cdot \vec{Y} )+\alpha \hat{e}_j (\nabla^j Y_i)X^i +\alpha \hat{e}_j (\nabla^j X_i)Y^i  \\ &=\underbrace{\nabla(\alpha) (\vec{X}\cdot \vec{Y})}_{I} +\underbrace{\alpha (\vec{X}\cdot\nabla)\vec{Y}}_{II}+ \underbrace{\alpha \vec{X}\times(\nabla\times\vec{Y})}_{III}  +\underbrace{\alpha \hat{e}_j (\nabla^j X_i)Y^i}_{IV} \label{eq_vec_identity2}\end{align}
 Substituting $\alpha=\frac{\tau_{ee}}{\hbar D_{\vec{k}}}\frac{\partial{f^{\text{lab}}( \vec{x},\vec{k})}}{{\partial\varepsilon_{\vec{k}} }}$, $\vec{X}=\bigl(\hbar \nabla_{ \vec{x}} \vec{u}({ \vec{x}}) \bigr)\vec{k}$ and  $\vec{Y}= \nabla_{\vec{k}} \varepsilon_{\vec{k}}$ we realize that the third term is identically zero as $\nabla_{\vec{k}}\times(\nabla_{\vec{k}} \varepsilon_{\vec{k}})=0 $.  The contribution from the four terms in RHS of the above identity can be written as: 
 \begin{align}\nonumber
g_{2}( \vec{x},\vec{k}) =&\frac{e\tau_{ee}}{\hbar^3 D_{\vec{k}}}\left(\left(\nabla_{\vec{k}}\varepsilon_{\vec{k}}\times \vec{B}\right)\cdot\frac{\partial \left(\frac{\tau_{ee}}{D_{\vec{k}}}\frac{\partial{f^{\text{lab}}( \vec{x},\vec{k})}}{{\partial\varepsilon_{\vec{k}} }}\right) }{\partial \vec{k}}\right)\bigl(\nabla_{\vec{k}}\varepsilon_{\vec{k}}\cdot \bigl(\bigl(\hbar \nabla_{ \vec{x}} \vec{u}({ \vec{x}}) \bigr)\vec{k}\bigr)\bigr) \\\nonumber&+\frac{e\tau_{ee}}{\hbar^3 D_{\vec{k}}}\left(\frac{\tau_{ee}}{D_{\vec{k}}}\frac{\partial{f^{\text{lab}}( \vec{x},\vec{k})}}{{\partial\varepsilon_{\vec{k}} }}\right)\left[\left(\nabla_{\vec{k}}\varepsilon_{\vec{k}}\times \vec{B}\right)\cdot\{\bigl(\bigl((\hbar\nabla_{ \vec{x}} \vec{u}({ \vec{x}}) )\cdot \vec{k}\bigr)\cdot\nabla_{\vec{k}} \vec{\bigr)}\nabla_{\vec{k}}\varepsilon_{\vec{k}}\}\right]\\&+\text{contribution from fourth term}
\label{eq_Start_S_U}\end{align}
Notice that the first two terms in RHS of the above equation have a similar pattern to the RHS in equation \ref{eq_g_2_S_E_mu}.
Now we look at the contribution from the fourth term of eq. \ref{eq_vec_identity2}. We write in Einstein summation convention(with the understanding that $\vec{k}$ and $ \vec{x}$ denote momentum and position and not the indices) and use $\epsilon$ for the Levi-civita tensor.
\begin{align}\nonumber \frac{e\tau_{ee}}{\hbar^2D_{\vec{k}}}\alpha\left(\nabla_{\vec{k}}\varepsilon_{\vec{k}}\times \vec{B}\right)_i  (\nabla^i X_p)Y^p=&\frac{e\tau_{ee}}{\hbar^2D_{\vec{k}}}\left(\frac{\tau_{ee}}{D_{\vec{k}}}\frac{\partial{f^{\text{lab}}( \vec{x},\vec{k})}}{{\partial\varepsilon_{\vec{k}} }}\right)\left(\nabla_{\vec{k}}\varepsilon_{\vec{k}}\times \vec{B}\right)_i \frac{\partial}{\partial k_i}\left(\frac{\partial}{\partial x_p}\bigl(u( \vec{x})^qk_q\bigr)\right)\frac{\partial \varepsilon_{\vec{k}}}{\partial k^p}\\ \nonumber =& \frac{e\tau_{ee}}{\hbar^2D_{\vec{k}}}\left(\frac{\tau_{ee}}{D_{\vec{k}}}\frac{\partial{f^{\text{lab}}( \vec{x},\vec{k})}}{{\partial\varepsilon_{\vec{k}} }}\right)\left(\epsilon_{imn}\frac{\partial\varepsilon_{\vec{k}}}{\partial k_m} B^n\right)\left(\frac{\partial}{\partial x_p}\bigl(u( \vec{x})^q\delta_q^i\bigr)\right)\frac{\partial \varepsilon_{\vec{k}}}{\partial k^p} \\ \nonumber =& \frac{e\tau_{ee}}{\hbar^2D_{\vec{k}}}\left(\frac{\tau_{ee}}{D_{\vec{k}}}\frac{\partial{f^{\text{lab}}( \vec{x},\vec{k})}}{{\partial\varepsilon_{\vec{k}} }}\right)\left(\epsilon_{imn}\frac{\partial\varepsilon_{\vec{k}}}{\partial k_m} B^n\right)\left(\frac{\partial}{\partial x_p}\bigl(u( \vec{x})^i\bigr)\right)\frac{\partial \varepsilon_{\vec{k}}}{\partial k^p} \\ \nonumber =& \frac{e\tau_{ee}}{\hbar^2D_{\vec{k}}}\left(\frac{\tau_{ee}}{D_{\vec{k}}}\frac{\partial{f^{\text{lab}}( \vec{x},\vec{k})}}{{\partial\varepsilon_{\vec{k}} }}\right)\left(\epsilon_{imn}\frac{\partial\varepsilon_{\vec{k}}}{\partial k_m} B^n\right)\left( \nabla_{\vec{k}} \varepsilon_{\vec{k}}\cdot\nabla_{ \vec{x}}\right)u( \vec{x})^i \\ \nonumber =& \frac{e\tau_{ee}}{\hbar^2D_{\vec{k}}}\left(\frac{\tau_{ee}}{D_{\vec{k}}}\frac{\partial{f^{\text{lab}}( \vec{x},\vec{k})}}{{\partial\varepsilon_{\vec{k}} }}\right)\left(\epsilon_{imn}\frac{\partial\varepsilon_{\vec{k}}}{\partial k_m}\frac{\partial \varepsilon_{\vec{k}}}{\partial k_p} B^n\right)\frac{\partial u( \vec{x})^i}{\partial x_p}\\ =& e\left(\frac{\tau_{ee}^2}{\hbar^2D_{\vec{k}}^2}\frac{\partial{f^{\text{lab}}( \vec{x},\vec{k})}}{{\partial\varepsilon_{\vec{k}} }}\right)\left(\epsilon_{imn}\frac{\partial\varepsilon_{\vec{k}}}{\partial k_m}\frac{\partial \varepsilon_{\vec{k}}}{\partial k_p} B^n\right)\frac{\partial u( \vec{x})^i}{\partial x_p}
\label{eq_grad_k_extra_u}
\end{align}
 Next, adding the contributions from equations \ref{eq_g_2_S_E_mu} and \ref{eq_Start_S_U} and substituting for the fourth term from equation \ref{eq_grad_k_extra_u}, we get
\begin{align} \nonumber &
 g_2( \vec{x},\vec{k}) =\frac{e\tau_{ee}}{\hbar^3D_{\vec{k}}}\left(\left(\nabla_{\vec{k}}\varepsilon_{\vec{k}}\times \vec{B}\right)\cdot\frac{\partial \left(\frac{\tau_{ee}}{D_{\vec{k}}}\frac{\partial{f^{\text{lab}}( \vec{x},\vec{k})}}{{\partial\varepsilon_{\vec{k}} }}\right) }{\partial \vec{k}}\right)(\nabla_{\vec{k}}\varepsilon_{\vec{k}}\cdot \vec{S}) \\ &+\frac{e\tau_{ee}^2}{\hbar^3D_{\vec{k}}^2}\left(\frac{\partial{f^{\text{lab}}( \vec{x},\vec{k})}}{{\partial\varepsilon_{\vec{k}} }}\right)\left(\left(\nabla_{\vec{k}}\varepsilon_{\vec{k}}\times \vec{B}\right)\cdot((\vec{S}\cdot\nabla_{\vec{k}} )\nabla_{\vec{k}}\varepsilon_{\vec{k}})\right)
+\frac{e\tau_{ee}^2}{\hbar^2D_{\vec{k}}^2}\frac{\partial{f^{\text{lab}}( \vec{x},\vec{k})}}{{\partial\varepsilon_{\vec{k}} }}\epsilon_{imn}\frac{\partial\varepsilon_{\vec{k}}}{\partial k_m}\frac{\partial \varepsilon_{\vec{k}}}{\partial k_p} B^n\frac{\partial u( \vec{x})^i}{\partial x^p} .
 \label{eq_g_2_S_final_app}\end{align}

\subsection*{A4. Demonstrating Onsager's reciprocity relations for transport coefficients up to $\tau_{ee}^2$ } 
\label{Appendix_Onsager_second}
In this appendix, we show that Onsager's relations hold for transport coefficients found in the main text.
In equation \ref{eq_L7_second_2d}, if exchange the indices $i$ and $l$ we get,
\begin{align}
  \nonumber 
  L^{7}_{lj,i}(\vec{B};\{\vec{\Omega_k}\}, \{\varepsilon_{\vec{k}}\} )=\frac{e^2}{\hbar^3}\int[d\vec{k}]k_l\frac{\partial \varepsilon_{\vec{k}}}{\partial k_j}&\left[
 \frac{\tau_{ee}^2}{D_{\vec{k}}} \epsilon_{pqr} \frac{\partial \varepsilon_{\vec{k}}}{\partial k_q}\pdv{\left(\frac{1}{D_{\vec{k}}}\right)}{k_p}\frac{\partial{f^{0}}}{{\partial\varepsilon_{\vec{k}} }}\pdv{\varepsilon_{\vec{k}}}{k_i}   \right. \\ &+ \left. \frac{\tau_{ee}^2}{D_{\vec{k}}^2}\frac{\partial{f^{0}}}{{\partial\varepsilon_{\vec{k}}}}\epsilon_{pqr} \frac{\partial \varepsilon_{\vec{k}}}{\partial k_q}\pdv{\varepsilon_{\vec{k}}}{k_i}{k_p} \right]B^r
\end{align}
We make the following transformation  $\vec{\Omega_k}\rightarrow -\vec{\Omega_{-k}}$, $\varepsilon_{\vec{k}}\rightarrow \varepsilon_{-\vec{k}}$ and  $\vec{B}\rightarrow-\vec{B}$ and change variable of integration $\vec{k}\rightarrow-\vec{k}$.
\begin{align}
  \nonumber 
  L^{7}_{lj,i}(-\vec{B};\{-\vec{\Omega_{-k}}\}, \{\varepsilon_{\vec{-k}}\} )=\frac{e^2}{\hbar^3}\int[d\vec{k}]k_l\frac{\partial \varepsilon_{\vec{k}}}{\partial k_j}&\left[
 \frac{\tau_{ee}^2}{D_{\vec{k}}} \epsilon_{pqr} \frac{\partial \varepsilon_{\vec{k}}}{\partial k_q}\pdv{\left(\frac{1}{D_{\vec{k}}}\right)}{k_p}\frac{\partial{f^{0}}}{{\partial\varepsilon_{\vec{k}} }}\pdv{\varepsilon_{\vec{k}}}{k_i}   \right. \\ &+ \left. \frac{\tau_{ee}^2}{D_{\vec{k}}^2}\frac{\partial{f^{0}}}{{\partial\varepsilon_{\vec{k}}}}\epsilon_{pqr} \frac{\partial \varepsilon_{\vec{k}}}{\partial k_q}\pdv{\varepsilon_{\vec{k}}}{k_i}{k_p} \right]B^r
\end{align}
Integrating by parts over $k_p$ in the second term and using the fact that $f^0\rightarrow 0$ as $k\rightarrow\infty$.

\begin{align}
  \nonumber 
  L^{7}_{lj,i}(-\vec{B};\{-\vec{\Omega_{-k}}\}, \{\varepsilon_{\vec{-k}}\} )=\frac{e^2}{\hbar^3}\int[d\vec{k}]&\left[k_l\frac{\partial \varepsilon_{\vec{k}}}{\partial k_j}
 \frac{\tau_{ee}^2}{D_{\vec{k}}} \epsilon_{pqr} \frac{\partial \varepsilon_{\vec{k}}}{\partial k_q}\pdv{\left(\frac{1}{D_{\vec{k}}}\right)}{k_p}\frac{\partial{f^{0}}}{{\partial\varepsilon_{\vec{k}} }}\pdv{\varepsilon_{\vec{k}}}{k_i}   \right. \\\nonumber &- \left.\delta_{pl}\frac{\partial \varepsilon_{\vec{k}}}{\partial k_j} \frac{\tau_{ee}^2}{D_{\vec{k}}^2}\frac{\partial{f^{0}}}{{\partial\varepsilon_{\vec{k}}}}\epsilon_{pqr} \frac{\partial \varepsilon_{\vec{k}}}{\partial k_q}\pdv{\varepsilon_{\vec{k}}}{k_i}  \right. \\\nonumber &- \left.k_l\frac{\partial ^2\varepsilon_{\vec{k}}}{\partial k_p\partial k_j} \frac{\tau_{ee}^2}{D_{\vec{k}}^2}\frac{\partial{f^{0}}}{{\partial\varepsilon_{\vec{k}}}}\epsilon_{pqr} \frac{\partial \varepsilon_{\vec{k}}}{\partial k_q}\pdv{\varepsilon_{\vec{k}}}{k_i} \right. \\\nonumber &- \left.\underbrace{2k_l\frac{\partial \varepsilon_{\vec{k}}}{\partial k_j} \pdv{(\tau_{ee})}{\varepsilon_{\vec{k}}}\pdv{\varepsilon_{\vec{k}}}{k_p}\frac{1}{D_{\vec{k}}^2}\frac{\partial{f^{0}}}{{\partial\varepsilon_{\vec{k}}}}\epsilon_{pqr} \frac{\partial \varepsilon_{\vec{k}}}{\partial k_q}\pdv{\varepsilon_{\vec{k}}}{k_i}}_{\text{vanishes because of being anti-symmetric in $p$ and $q$}}\right. \\\nonumber &-2\left.k_l\frac{\partial \varepsilon_{\vec{k}}}{\partial k_j} \frac{\tau_{ee}^2}{D_{\vec{k}}}\pdv{\left(\frac{1}{D_{\vec{k}}}\right)}{k_p}\frac{\partial{f^{0}}}{{\partial\varepsilon_{\vec{k}}}}\epsilon_{pqr} \frac{\partial \varepsilon_{\vec{k}}}{\partial k_q}\pdv{\varepsilon_{\vec{k}}}{k_i} \right. \\\nonumber &- \left.\underbrace{k_l\frac{\partial \varepsilon_{\vec{k}}}{\partial k_j} \tau_{ee}^2\frac{1}{D_{\vec{k}}^2}\frac{\partial^2{f^{0}}}{{\partial\varepsilon_{\vec{k}}^2}}\pdv{\varepsilon_{\vec{k}}}{k_p}\epsilon_{pqr} \frac{\partial \varepsilon_{\vec{k}}}{\partial k_q}\pdv{\varepsilon_{\vec{k}}}{k_i}}_{\text{vanishes because of being anti-symmetric in $p$ and $q$}}
 \right. \\ &- \left.\underbrace{k_l\frac{\partial \varepsilon_{\vec{k}}}{\partial k_j} \tau_{ee}^2\frac{1}{D_{\vec{k}}^2}\frac{\partial{f^{0}}}{{\partial\varepsilon_{\vec{k}}}}\epsilon_{pqr} \frac{\partial^2 \varepsilon_{\vec{k}}}{\partial k_p\partial k_q}\pdv{\varepsilon_{\vec{k}}}{k_i}}_{\text{vanishes because of being anti-symmetric in $p$ and $q$}}\right]B^r
 \label{eq_Onsager_demonstrate}
\end{align}
Therefore,

\begin{align}
  \nonumber 
  L^{7}_{lj,i}(-\vec{B};\{-\vec{\Omega_{-k}}\}, \{\varepsilon_{\vec{-k}}\} )=\frac{e^2}{\hbar^3}\int[d\vec{k}]&\left[-k_l\frac{\partial \varepsilon_{\vec{k}}}{\partial k_j}
 \frac{\tau_{ee}^2}{D_{\vec{k}}} \epsilon_{pqr} \frac{\partial \varepsilon_{\vec{k}}}{\partial k_q}\pdv{\left(\frac{1}{D_{\vec{k}}}\right)}{k_p}\frac{\partial{f^{0}}}{{\partial\varepsilon_{\vec{k}} }}\pdv{\varepsilon_{\vec{k}}}{k_i}   \right. \\\nonumber &- \left.\frac{\partial \varepsilon_{\vec{k}}}{\partial k_j} \frac{\tau_{ee}^2}{D_{\vec{k}}^2}\frac{\partial{f^{0}}}{{\partial\varepsilon_{\vec{k}}}}\epsilon_{lqr} \frac{\partial \varepsilon_{\vec{k}}}{\partial k_q}\pdv{\varepsilon_{\vec{k}}}{k_i}  \right. \\ &- \left.k_l\frac{\partial ^2\varepsilon_{\vec{k}}}{\partial k_p\partial k_j} \frac{\tau_{ee}^2}{D_{\vec{k}}^2}\frac{\partial{f^{0}}}{{\partial\varepsilon_{\vec{k}}}}\epsilon_{pqr} \frac{\partial \varepsilon_{\vec{k}}}{\partial k_q}\pdv{\varepsilon_{\vec{k}}}{k_i}\right]B^r
 \label{eq_L_7_mod}
\end{align}
Comparing equations \ref{eq_L_7_mod} and \ref{eq_L3_second_2d}, we conclude that the Onsager relation in equation \ref{eq_Onsager_relatiosA} is satisfied. The off-diagonal Onsager relation for $L^6$ and $L^8$ and the diagonal relation for $L^9$ can be similarly shown. 

\bibliography{References}

\end{document}